\documentstyle[preprint,tighten,aps,epsf,floats,aps10]{revtex}

\begin{document}
\date{\today, submitted to J. Phys. Condens. Matter}
\draft
\preprint{}

\title{Pseudogap State in High-$T_c$ Superconductors: an 
Infrared Study}

\author{A.V.~Puchkov$^{(1)}$, D.N.~Basov$^{(2)}$, and T.~Timusk$^{(1)}$} 

\address{ $^{(1)}$ Department of Physics and Astronomy, 
McMaster University, Hamilton, Ontario, Canada L8S 4M1}

\address{ $^{(2)}$ Department of Physics, Brookhaven National 
Laboratory, Upton, New York, 11973-5000}
\maketitle

\begin{abstract}
We report on a study of the electromagnetic response of 
three different families of high-$T_c$ superconductors that in 
combination allowed us to cover the whole doping range from under- to 
overdoped. The discussion is focused on the $ab$-plane 
charge dynamics in the {\it pseudogap state} which is realized in 
underdoped materials below a characteristic temperature $T^*$; a 
temperature that can significantly exceed the superconducting 
transition temperature $T_c$. We explore the evolution of the 
pseudogap response by changing the doping level, by varying the
temperature from the above to below $T^*$, or by introducing 
impurities in the underdoped compounds. We employ a memory-function 
analysis of the $ab$-plane optical data that allows us to observe the 
effect of the pseudogap most clearly. We compare the infrared data 
with other experimental results, including $c$-axis optical 
response, dc transport, and angular resolved photoemission. 
\end{abstract} 


\section{INTRODUCTION}
\begin{figure}[htpb]
\leavevmode 
\epsfxsize=0.5\columnwidth
\centerline{\epsffile{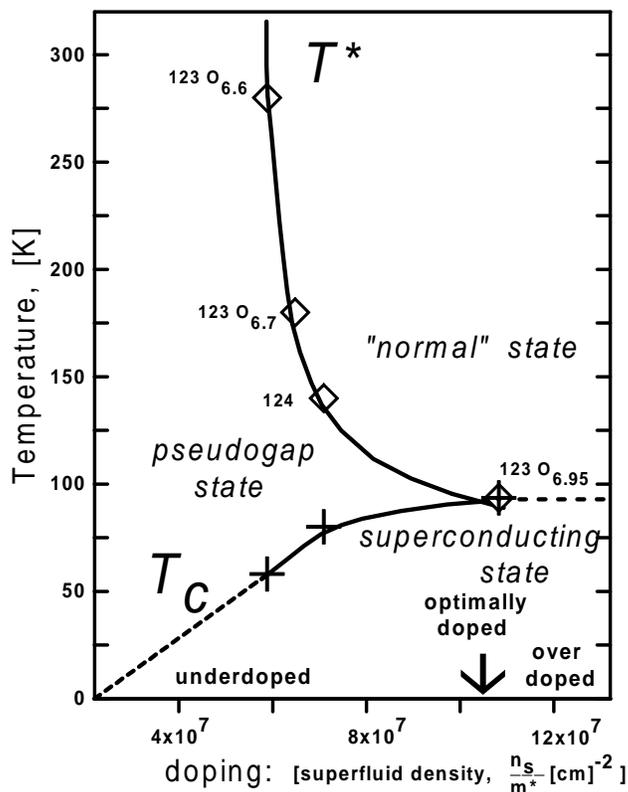}} 
\caption{Schematic phase diagram of the cuprate superconductors. In 
the underdoped regime a pseudogap state forms below a temperature 
$T^*>T_c$. The curves for $T^*$ and $T_c$ cross at optimal doping 
where the pseudogap and the superconducting gap develop at the same 
temperature. $T^*$ is determined from the $c$-axis conductivity and 
the doping level from the superfluid density 
$\omega_{ps}^2 = n_s/m^*$ in the CuO$_2$ planes.
}
\label{phase-dia1}
\end{figure}

\setcounter{page}{1}
There is mounting evidence that the normal state of underdoped 
high-$T_c$ superconductors (HTSC) is dominated by a pseudogap. A 
number of physical probes show that below a characteristic 
temperature $T^*$, which can be well above the superconducting 
transition temperature $T_c$, the physical response of HTSC 
materials can be interpreted in terms of the formation of a partial 
gap or a pseudogap by which we mean a suppression of the density of 
low-energy excitations. This gap persists in the superconducting 
state. $T^*$ decreases with increasing doping in the underdoped 
regime and since $T_c$ rises with doping, the two curves meet at the 
optimal doping level, as shown in the schematic phase diagram in 
Fig.~\ref{phase-dia1}. 

The earliest experiments to reveal gap-like behavior in the 
normal state were nuclear magnetic resonance (NMR) measurements 
of the Knight shift,\cite{warren89,yoshinari} which probes the 
uniform spin susceptibility. In conventional superconductors and the 
cuprates at optimal doping, the Knight shift is temperature 
independent in the normal state but drops rapidly below $T_c$ due to 
pairing of electronic spins into (singlet) superconducting Cooper 
pairs. In underdoped cuprates, however, the Knight shift begins to 
drop well above the superconducting transition temperature. Warren 
{\it et al.} concluded that in these materials spin pairing takes 
place well above the bulk superconducting transition at $T_c$, 
thus producing a normal-state energy gap, referred to as a "spin 
gap".\cite{warren89} 

\vspace{0.3in}

Deviations from the well known linear temperature dependence of 
the $ab$-plane resistivity,\cite{gurvitch} $\rho_{ab}(T)$, were 
observed in underdoped cuprates as 
well,\cite{bucher93,batlogg,ito93,walkes93} with the slope of 
$\rho_{ab}(T)$ changing below a characteristic temperature $T^*$. 
As the doping is increased towards the optimal level, $T^*$ 
decreases and the near-optimal doping $\rho_{ab}(T)$ is linear over
the range of temperatures from $T_c$ to above
800~K.\cite{ito93,batlogg} 

The magnitude of $T^*$ as well as its variation with doping suggest 
that the suppression of the spin susceptibility observed in NMR 
measurements and the change of slope of $\rho_{ab}(T)$ have a 
common physical origin. It has been suggested that if the 
scattering responsible for the linear temperature dependence 
of ${\rho_{ab}}(T)$ involves scattering on spin fluctuations, then 
the spin gap seen in NMR below $T^*$ would naturally account for the 
depression of $\rho_{ab}(T)$ below $T^*$ as well. Similar evidence 
for the suppression of the spin susceptibility has been extracted 
from neutron scattering experimental results.\cite{rossat-mignod91} 
Specific heat measurements on underdoped YBa$_2$Cu$_3$O$_x$ (Y123), 
however, show that there is a large decrease in entropy below a 
temperature, closely related to the $T^*$, which can not be 
accounted for by assuming that a gap in the spin degrees of 
freedom is solely responsible.\cite{loram94} 

There is spectroscopic evidence of anomalies in the
properties of HTSC that were originally associated with the formation 
of the superconducting gap, but were found to occur at $T>T_c$ in 
underdoped samples. The shift in the position and width of Raman 
frequencies of certain phonons, associated with the onset of 
superconductivity,\cite{friedl90b} were shown to occur in the normal 
state of underdoped cuprates and it was suggested they were related 
to the spin gap.\cite{litvinchuk92} Similarly, broad peaks in the 
electronic Raman continuum, also interpreted as an evidence for the 
formation of a superconducting gap,\cite{cooper88} were found to 
occur well above $T_c$ in underdoped samples.\cite{slakey90} 

Indications of normal-state, gap-like anomalies in underdoped 
cuprates were observed in infrared optical measurements as well. 
To a first approximation the $ab$-plane optical properties of HTSC are those 
of a metal where the charge carriers move coherently through the 
lattice. Such coherent motion gives rise to a conductivity peak, 
centered at zero frequency, called the Drude peak, whose width is a measure 
of the inverse lifetime of the carriers. In this paper we call 
systems that have a conductivity peak at zero frequency coherent 
systems.  In contrast, transport in the c-axis direction does not 
show a peak a zero frequency and we call this incoherent transport. 
On  closer examination the reflectance of most high 
temperature superconductors was found to deviate from simple Drude 
behavior which predicts a reflectance decreasing  monotonically with 
frequency. A structure in the form of a "knee" was found at 
approximately 500~cm$^{-1}$. This structure was sometimes interpreted 
as a manifestation of a conventional superconducting gap. It has been 
found, however, that in underdoped materials the knee starts to 
develop already in the normal state. 
\cite{reedyk88a,thomas88a,orenstein90,kamaras90,vandermarel91a,schlesinger90,rotter91,schlesinger94} 
A comparison with other probes suggests that the knee structure and 
deviations observed in the dc transport and NMR experiments all occur 
at a characteristic temperature remarkably similar to $T^*$. The 
corresponding changes in the complex optical conductivity 
$\sigma(\omega)=\sigma_1(\omega) + i \sigma_2(\omega)$ involve a 
shift of part of the $\sigma_1(\omega)$ spectral weight from 300-
700~cm$^{-1}$ to {\it lower} frequencies, resulting in a marked 
narrowing of the Drude peak. This behavior is in accord with decreasing dc 
resistivity and was interpreted in terms of coupling of electrons to 
the longitudinal optical (LO) phonons\cite{timusk91,reedyk92e} or as 
a manifestation of the spin 
gap.\cite{orenstein90,rotter91,schlesinger94} 

It should be emphasized that in the case of a {\it coherent} system, 
such as the underdoped cuprates in $ab$-plane direction, there is 
no direct mapping between the electronic density of states (DOS) and 
the shape of the real part of conductivity, $\sigma_1(\omega)$. For 
example, even if there is a gap in electronic DOS and its magnitude 
is larger than the characteristic energy associated with the {\it 
elastic} scattering (clean limit\cite{kamaras90,timusk88b}), the gap 
will not manifest itself in the $\sigma_1(\omega)$ spectra. In the 
same way, a pseudogap in the electronic DOS of a coherent system, 
that may appear due to strong interactions in the system, does not 
appear as an obvious gap in the conductivity.

\begin{figure}[htpb]
\leavevmode
\epsfxsize=0.55\columnwidth
\centerline{\epsffile{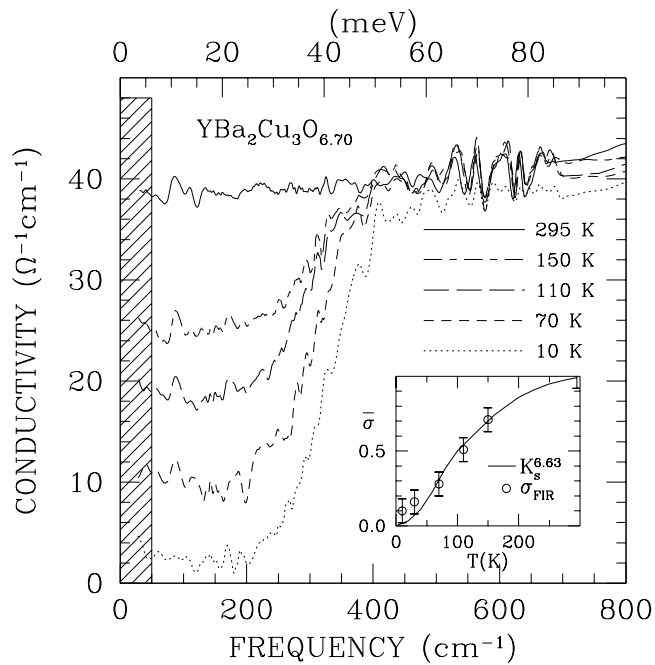}
 }
 \caption{The c-axis conductivity of an underdoped Y123 crystal. The c- 
axis conductivity is temperature and frequency independent for 
$T>T^*$ but develops a marked gap-like depression below $T^*$. As 
the temperature is lowered the pseudogap deepens. The inset: The NMR Knight 
shift (normalized at 300~K) is plotted as a function of temperature 
for an underdoped Y123 crystal. The circles show the low frequency 
$c$-axis conductivity for samples of the same doping level. The 
curves suggest that the Knight shift, a conventional measure of the 
density of states at the Fermi level, and the $c$-axis conductivity 
are depressed by the same process in the pseudogap state.}
\label{homes-caxis}
\end{figure}

The charge dynamics along the interplane $c$-direction is {\it 
incoherent}, at least in the underdoped materials. While both
underdoped Y123 and YBa$_2$Cu$_4$O$_8$ (Y124) compounds, collectively 
referred as YBCO, exhibit a gap-like depression in the c-axis 
conductivity as well,\cite{homes93,basov94c,tajima95} the $c$-axis 
conductivity shows no coherent peak at low frequencies, 
Fig.~\ref{homes-caxis}. 
Contrary to the $ab$-plane response, as the temperature decreases 
from 300~K to $T_c$ the $c$-axis $\sigma_1(\omega)$ spectral weight 
is transferred from the gap region to {\it higher} 
frequencies.\cite{homes93,leggett92} This is inferred from the sum 
rule for the optical conductivity, or spectral weight. The spectral 
weight lost at low frequency, as the gap develops, does not go to low 
frequencies since the magnitude of the 
low-frequency conductivity is in good agreement with the dc 
resistivity\cite{homes95} which shows a "semiconducting" behavior 
(i.e. resistivity increases at low T). Thus by default it must go to  
high frequencies and in the the case of La$_{2-x}$Sr$_x$CuO$_4$ (La214)
the spectral weight has been shown to transfer to the 1 eV 
region.\cite{basov95c} 
A number of mechanisms have 
been proposed that would result in an incoherent conductivity 
spectrum. 
\cite{kumar90,kumar92,rojo93,graf93,ioffe93,nyhus95,clarke95,alexandrov96} 
The c-axis conductivity depression in both Y123 and Y124 occurs at a 
temperature scale that matches the spin susceptibility determined 
from the NMR measurements. This is shown in the inset of 
Fig.~\ref{homes-caxis} where the Knight shift\cite{takigawa91} is 
shown along with the experimental c-axis conductivity. The onset 
energy of the c-axis gap in YBCO is ${\approx}200$~cm$^{-1}$ and the 
half value point is ${\approx}300$~cm$^{-1}$. Above 500~cm$^{-1}$, 
the c-axis conductivity is both temperature and frequency 
independent.\cite{homes95} 

A pseudogap has also been observed in the c-axis conductivity of 
La214 where for x=0.14 a very large gap has 
been reported \cite{basov95c} and for x=0.12 a gap of the same 
magnitude as in YBCO can clearly be seen.\cite{uchida96} The 
Pb$_2$Sr$_2$(Y/Ca)Cu$_2$O$_8$ material also shows a c-axis 
pseudogap.\cite{reedyk96} 

Recent angular-resolved photoemission (ARPES) results for
Bi$_2$Sr$_2$CaCu$_2$O$_{8+\delta}$ (Bi2212) show evidence of a 
normal-state gap-like depression of the electronic density of states 
of underdoped cuprates as 
well.\cite{marshall96,loeser96,ding96} The momentum dependence of 
this gap resembles that of the $d_{x^2- y^2}$ gap observed in the 
superconducting state.\cite{loeser96} This, and the fact that no 
significant changes are observed upon crossing into the 
superconducting regime, have led to the suggestion that the 
normal-state gap may be a precursor of the superconducting gap. As 
the doping level is increased to near- and above optimal the 
normal-state gap-like feature disappears. 

In the following, we summarize the recent experimental optical 
results obtained from several series of HTSC materials at doping 
levels ranging from underdoped to strongly overdoped. We find that 
in the pseudogap state the optical response of underdoped cuprates is 
marked by an increase in coherence of the electronic system. 
Since the coherence effects are seen more clearly through the 
frequency dependent charge-carrier scattering or memory function 
analysis, we have chosen to use this approach. The essential features 
of this very general formalism are described in Section III. We will 
also restrict our survey to the ab-plane optical properties. The 
c-axis optical data are less complete since the large, thick crystals 
needed for this work are not available for all systems. We will, 
however, try to address the question of the correlation between the 
$ab$-plane and the $c$-axis pseudogap properties as we change materials, 
doping and temperature. 

\section{THE EXPERIMENTAL TECHNIQUE}

In order to cover a broad range of doping regimes we performed 
reflectivity measurements upon three families of high-$T_c$ cuprates: 
YBCO (including Y123 and Y124), Bi2212
and Tl$_2$Sr$_2$CuO$_{6+\delta}$ (Tl2201). This was necessary 
because, with the exception of La214, none of the existing cuprates 
allows one to explore a full spectrum of doping regimes. For 
instance, Y123 materials can be conveniently underdoped by reducing 
the amount of oxygen from the optimally doped level at $x=6.95$. 
However these crystals are not suitable for strong overdoping. On the 
contrary, Tl2201 samples could be only overdoped by introducing 
interstitial O atoms between the TlO planes so that $T_c$ is 
suppressed from about 90 K in the {\it stoichiometric} composition 
down to less than 4 K in the overdoped composition. Single crystals 
of Bi2212 can be both overdoped and underdoped, but the suppression 
of the critical temperature is very limited on the overdoped side. We 
have also used Bi2212 crystals with 20\% of Bi substituted by Pb 
which allows one to achieve a higher degree of overdoping. The 
influence of disorder on the infrared response of YBCO crystals was 
studied by substituting Cu atoms in the CuO$_2$ planes with 
Zn.\cite{dabrowskizn}

The response of YBCO crystals was studied in three carrier density 
regimes: in an optimally doped Y123 crystal with oxygen content set 
at x=6.95 ($T_c$=93.5~K), in the {\it same} crystal deoxygenated down 
to x=6.6 ($T_c$=59~K)\cite{liang} and in a double-chained Y124 
crystal with $T_c$=82~K.\cite{dabrowski} The carrier density in the 
stoichiometric and naturally untwinned Y124 corresponds to that of 
Y123 samples with $x{\simeq}6.85$. The Y123 crystal was mechanically 
detwinned so that both $a$- and $b$-axis components of the 
conductivity tensor were obtained independently, allowing us to probe 
the response of the CuO$_2$ planes {\it only} without any 
contribution from the charge reservoir structural blocks. 

We have also performed reflectivity measurements on two underdoped 
($T_c$=67~K and $T_c$=82~K), one optimally doped 
($T_c=90$~K), and one overdoped ($T_c$=82~K) Bi2212 single 
crystal. All the Bi2212 crystals were prepared from the as-grown 
crystals by annealing in argon and/or oxygen.\cite{Pat} To achieve a 
higher degree of overdoping we have performed measurements on 
Pb-doped Bi2212 Bi$_{1.66}$Pb$_{0.34}$Sr$_2$CaCu$_2$O$_{8+\delta}$ 
((Bi/Pb)2212) with $T_c$=70~K (optimum $T_c$=88~K). The two T2201 
single crystals used in the measurements had $T_c$'s of 90~K (highest 
$T_c$ achievable) and 23~K (strongly overdoped). The superconducting 
transition temperatures for all samples were measured in a SQUID 
magnetometer.

The reflectivity was measured over a broad energy range: from 30 - 
50~cm$^{-1}$ up to 20,000~cm$^{-1}$ for Y123, Y124, and Bi2212 
samples; from 30 - 50 cm$^{-1}$ up to 50,000~cm$^{-1}$ for Tl2201 
samples. The far-infrared reflectance measurements were carried out 
using a rapid scan interferometer with focused optics on a sample 
mounted in a continuous flow cryostat. For near-infrared and visible 
measurements, a grating spectrometer with appropriate detector-filter 
combinations with overlapping frequency ranges was used. To obtain 
the absolute value of the reflectance, geometrical scattering losses 
were accounted for by {\it in situ} evaporation of a metallic film 
(Au or Al) onto the surface of the sample. The coated sample was then 
remeasured and the absolute value of the reflectance R is then given 
by the ratio of spectra before and after plating, corrected for the 
absolute reflectance of the metallic film.\cite{homes93a}

The complex optical conductivity $\sigma_1(\omega) + 
i\sigma_2(\omega)$ of single crystalline samples was obtained from 
Kramers-Kronig analysis of reflectivity. To perform the required 
integrations it was necessary to extend the reflectance beyond the 
actually measured range.\cite{timusk1} Below the lowest frequency 
measured we have tried different types of reflectivity 
approximations, from the Hagen-Rubens formula to a straight line 
between unity at zero frequency and the last experimental point. We 
found that in the frequency region that we will be of interest in 
this work ($\omega>100$~cm$^{-1}$) the particular choice of the 
low-frequency approximation is not important. At high frequencies, 
the reflectivity of Y123 and Y124 samples was extended using the results 
of previous measurements\cite{romberg} up to 300,000~cm$^{-1}$. 
The reflectivity of Tl2201 was approximated by 
a constant between 50,000-200,000~cm$^{-1}$. For Bi2212, the results 
of ellipsometric measurements\cite{humlicek90} were used between 
20,000-50,000~cm$^{-1}$ while above this frequency range a constant 
reflectivity approximation was used up to 200,000~cm$^{-1}$. Above 
200,000~cm$^{-1}$ for Bi2212 and Tl2201 and above 300,000~cm$^{-1}$ 
for YBCO the reflectivity was allowed to fall as $\omega^{-4}$. In 
the case of the Y123 material we compared the optical constants 
obtained using Kramers-Kronig analysis with those obtained more 
directly by optical ellipsometry in 2-5 eV range,\cite{cooper93} and 
excellent agreement was found. This attests to the reliability of the 
results obtained through Kramers-Kronig analysis of the reflectance. 

\section{THE EXTENDED DRUDE FORMALISM.}

\subsection{ Complex Memory Function.}

The classical Drude formula for the dynamical conductivity 
$\sigma(\omega)=\sigma_1(\omega)+i\sigma_2(\omega)$ 
\cite{timusk1,Ashcroft} can be obtained by using a standard Boltzmann 
equation and approximating the collision integral with a single 
collision frequency $1/\tau$. The Drude formula describes the 
free-carrier contribution to $\sigma_1(\omega)$ as a Lorentzian 
peak centered at zero frequency with an oscillator strength 
$\omega_p^2/8$, where $\omega_p^2=[e^2/(3\pi^2\hbar)]{\int}{\bf v} 
{\cdot} d{\bf S}_F$ and ${\bf v}$ is the electron velocity and 
${\bf S}_F$ is the element of Fermi surface. For a spherical Fermi 
surface $\omega_p^2=4{\pi}ne^2/m_e$, where $n$ is the free-carrier 
density and $m_e$ is the electronic band mass. The Lorentzian width 
is determined by a constant scattering rate $1/\tau$. The imaginary 
part of $\sigma(\omega)$ is just the real part multiplied by 
$\omega\tau$:

\begin{equation}
\sigma(\omega)= {1 \over {4\pi}}
{{\omega_p^2} \over {1/\tau-i\omega}}={{\omega_p^2} \over {4\pi}} 
[{\tau \over {1+(\omega\tau)^2}} + i{{\tau^2\omega} \over 
{1+(\omega\tau)^2}}]
\end{equation}

A derivation of Eq.~1 by using the standard kinetic Boltzmann 
equation assumes that the elementary system excitations are 
well-defined. However, a description of a system by using elementary 
excitations is possible, strictly speaking, only if the 
(energy) width of the wave packet representing the electronic 
excitation is small compared to the energy of the packet. In more 
formal language, for the approximations leading to Eq.~1 to be valid, 
a spectral function of electronic excitations, defined as:

\begin{equation}
A({\bf k},\omega)=-{1 \over \pi}|ImG({\bf k},\omega)|={1 \over \pi}
{{Im \Sigma(\omega}) \over {(\omega-\epsilon_k-Re\Sigma(\omega))^2+
(Im\Sigma(\omega))^2}},
\end{equation}
must be a narrow peak centered at 
$\omega=\epsilon_k+Re\Sigma(\omega)$. Here $G({\bf k},\omega)$ is a 
Green function of electronic excitation and $\Sigma(\omega)$ is the 
self-energy part. The narrowness of the peak means that 
the excitation energy must be much larger than the damping term 
$\gamma(\omega)=-2Im\Sigma(\omega)$. This is certainly true in case 
of standard Fermi-liquid theory, where $Re\Sigma(\omega)\sim\omega$ 
and $Im\Sigma(\omega)\sim\omega^2$ so that the electronic excitations 
(quasiparticles) are well-defined at zero temperature and energies 
close to the Fermi energy $E_F$.\cite{AGD} It can also be 
shown\cite{Shulga} that a weak electron-phonon coupling, although it
violates the quasiparticle description at energies very close to 
$E_F$, does not drastically change the transport properties at low 
temperatures, since in this case the number of electronic states 
where the quasiparticle description is violated is small. Therefore, 
the Drude formula is applicable only for simple metals at 
low frequencies and low temperatures where elastic scattering from 
impurities and weak quasielastic scattering from thermally excited 
excitations such as phonons dominate.\cite{timusk1,Shulga} 

On the other hand, following the original ideas of
Anderson,\cite{anderson87} it is now widely accepted that the 
electronic system of HTSC materials represent a new kind of quantum 
liquid and the simple Fermi-liquid quasiparticle description
is not applicable to the normal-state properties of 
these materials. For example, the key ingredient of the 
phenomenological "marginal" Fermi-liquid theory,\cite{MFL} advanced 
to explain these properties, is the assumption that the 
$Im\Sigma(\omega)\sim\omega$ and, consequently, $Re\Sigma(\omega)$ 
diverges logarithmically at the Fermi energy, thus making $G({\bf 
k},\omega)$ entirely incoherent at $E_F$. On a more microscopic
level, a similar result is expected for the quasi-one dimensional 
Hubbard model, which was identified by Anderson as an appropriate 
paradigm for the resonant valence bond (RVB) 
description.\cite{anderson87} Even in more Fermi-liquid-like 
scenarios, sufficiently strong coupling of an electronic system to a 
bosonic energy spectrum may result in a violation of the 
quasiparticle description.\cite{Shulga} In addition, the 
Fermi energy is estimated to be only $E_F=1-2$~eV, which is not much 
larger than the energies probed in infrared experiments (4-300~meV). 
Such a low $E_F$ may be another reason for violation of a 
quasiparticle description. Since this implies the absence of 
well-defined elementary excitations, the approximations used to 
obtain Eq.~1 are not justified. The breakdown of the quasiparticle 
description has also been discussed by Emery and Kivelson in the 
context of abnormally short values of the mean free path that lead to 
the violation of the Ioffe-Regel criterion.\cite{emeryprl}

However, the optical conductivity can be described in a much more 
general way by making the damping term in the Drude formula complex 
and frequency-dependent: $1/\tau=M(\omega)=M'(\omega)+iM''(\omega)$, 
where $M(\omega)$ is called a memory 
function.\cite{Allen_M,Mori,Gotze} The $M(\omega)$ satisfies 
$M'(\omega)=M'(-\omega)$ and $M''(\omega)=-M''(-\omega)$. The 
consequences of this formalism, usually referred to as the extended 
Drude model, have been derived for the infrared conductivity of 
metals with a strong electron-phonon interaction by Allen\cite{Allen} 
and Allen and Mikkelsen\cite{Allen_M} for the case of zero 
temperature. The analysis was later extended for the case of finite 
temperatures by Shulga {\it et al.}\cite{Shulga} It is also believed 
that the resulting theory is valid in the case of coupling of a Fermi 
liquid to any bosonic energy spectrum. Some aspects of the extended 
Drude model were also examined in detail by G\"otze and 
W\"olfe.\cite{Gotze} We are not aware of any 
quantitative predictions regarding the extended Drude model in the 
completely non-Fermi-liquid scenario, such as the Luttinger-liquid 
theory. Therefore, in the following we will employ the Fermi-liquid 
terminology. The formalism has been previously applied to 
transition-metal compounds,\cite{Allen_M} heavy-fermion 
materials,\cite{Webb,Shulga_1} and the HTSC 
cuprates.\cite{thomas88a,collins90,rieck95,basov96c,puchkov96b}

Rewriting the complex conductivity $\sigma(\omega)$ in terms of a 
complex memory function, 
$M(\omega,T)=1/\tau(\omega,T)-i\omega\lambda(\omega,T)$, one 
obtains\cite{timusk1,Mori}:

\begin{equation}
\sigma(\omega,T)={1 \over {4\pi}} {{\omega_p^2} 
\over {M(\omega,T)-i\omega}}= 
{1 \over {4\pi}} {{\omega_p^2} 
\over {1/\tau(\omega,T)-i\omega[1+\lambda(\omega,T)]}}
\end{equation}

Although, in the case of a metal, Eq.~3 can be obtained using 
Boltzmann-equation formalism with a frequency dependent scattering 
rate,\cite{Allen_M} this form has in fact a range of validity more 
general than the Boltzmann-equation approach.\cite{Allen_M,Mori} 
Adopting the Boltzmann-type terminology, the quantities 
$1/\tau(\omega,T)$ and $\lambda(\omega,T)$ describe the 
frequency-dependent scattering rate and mass enhancement of 
electronic excitations due to many-body interactions. 

Using the more general form of Eq.~3, one can check the range of 
validity of the classical Drude formula of Eq.~1 by expanding the 
memory function into Taylor series for small frequencies:

\begin{equation}
\lim_{\omega\rightarrow 
0}M(\omega)={1 \over \tau(0)}-i\lambda(0)\omega + O(\omega^2)
\end{equation} 

Substituting this into Eq.~3 one finds:

\begin{equation}
\sigma(\omega,T)= {1 \over {4\pi}}
{{\omega_p^2} \over {1/\tau(0)-i\omega(1+\lambda(0))}},
\end{equation}
recovering Eq.~1. The classical Drude result is thus valid 
whenever expansion of Eq.~4 makes sense and $\lambda(0)$ is small 
compared to unity.

Eq.~3 can be reduced to the familiar Drude form of Eq.~1 by 
introducing the so called renormalized scattering rate 
$1/\tau^*(\omega,T)=1/[\tau(\omega,T)(1+\lambda(\omega,T))]$ and the 
effective plasma frequency 
$\omega_p^{*2}(\omega,T)=\omega_p^2/(1+\lambda(\omega,T))$: 

\begin{equation}
\sigma(\omega,T)={1 \over {4\pi}} {{\omega_p^{*2}(\omega,T)} 
\over {1/\tau^*(\omega,T)-i\omega}}
\end{equation}

As it can be seen from this equation, the optical conductivity is now 
composed of an infinite set of Drude peaks, each describing 
$\sigma(\omega)$ in the vicinity of a particular frequency $\omega$ 
with a set of parameters $1/\tau^*(\omega)$ and $\lambda(\omega)$ 
(for simplicity in the following we will drop the temperature 
parameter when it is not relevant to a discussion). The 
$1/\tau^*(\omega)$ has a phenomenological meaning of a width of the 
Drude peak local to a frequency $\omega$ while $\lambda(\omega)$ 
represents the interaction-induced velocity renormalization. The 
renormalized scattering rate $1/\tau^*(\omega)$ is not causal and, 
other than the local Drude width, does not have a real physical sense 
as it includes both the velocity renormalization and the lifetime 
effects.

On the other hand, $1/\tau(\omega)$ is, up to a constant, the real 
part of $1/\sigma(\omega)$

\begin{equation}
1/\tau(\omega)={\omega_p^2 \over {4\pi}} Re({1 \over \sigma(\omega)}),
\end{equation}
that is, a real part of a physical response function. In the limit of 
zero frequency the normal-state optical conductivity is completely 
real and Eq.~4 becomes 
$1/\sigma_{dc}(T)=\rho_{dc}(T)=m_e/(\tau(T)ne^2)$, where 
$\rho_{dc}(T)$ is the dc resistivity. This is the same form as the 
relaxation-time expression for the dc resistivity of a free electron 
gas and therefore $\tau(\omega,T)|_{\omega=0}$ may be viewed as an 
electronic lifetime. 

The mass enhancement factor $\lambda(\omega)$ is given as the 
imaginary part of $1/\sigma(\omega)$:

\begin{equation}
1+\lambda(\omega)=-{\omega_p^2 \over {4\pi}} {1 \over \omega} Im({1 
\over \sigma(\omega)}).
\end{equation}

The total plasma frequency $\omega_p^2$ in Eqs.~7,8 can be can be 
found from the sum rule 
$\int_0^{\infty}\sigma_1(\omega)d\omega=\omega_p^2/8$. Since 
$\sigma(\omega)$ is causal, $\lambda(\omega)$ and $1/\tau(\omega)$ 
are not independent and are related by the Kramers-Kronig 
relation.\cite{timusk1} Using the relations 
$1/\tau(\omega)=1/\tau(-\omega)$ and 
$\lambda(\omega)=\lambda(-\omega)$ we obtain:

\begin{equation}
\lambda(\omega)={2 \over \pi} {\cal P}\int_0^\infty {{1/\tau(\Omega)} 
\over {\Omega^2-\omega^2}} d\Omega
\end{equation}

\begin{equation}
1/\tau(\infty)-1/\tau(\omega)={2 \over \pi} {\cal P}\int_0^\infty 
{{\Omega^2\lambda (\Omega)} \over {\Omega^2-\omega^2}} d\Omega
\end{equation}

If $1/\tau(\omega)$ and $\lambda(\omega)$ have no poles at $\omega=0$ 
one immediately obtains the following useful relation:

\begin{equation}
1/\tau(\infty)-1/\tau(0)={2 \over \pi} \int_0^\infty \lambda(\Omega) 
d\Omega
\end{equation}

We see that the complex memory function $M(\omega)$ is a 
physical response function and experimental data can be presented in 
terms of $M(\omega,T)$ or the complex optical 
conductivity $\sigma(\omega,T)$ equally well. The particular choice 
should be made judging from the situation at hand. For example, the 
memory function analysis may be useful if one is interested in the 
relaxation processes that determine a system response to 
electromagnetic radiation. Also, in certain cases the memory 
function is easier to calculate analytically, thus making it easier 
to analyze the physics behind the system behavior using experimental 
results for $M(\omega)$. For example, simple analytical formulae for 
$M(\omega)$ have been derived for electron-phonon scattering while 
the optical conductivity has to be calculated 
numerically.\cite{Shulga}

Finally, we would like to stress that, although Eq.~3 is very 
general, obviously the {\it interpretation} of experimental results 
for $M(\omega,T)$ in terms of scattering rate and mass enhancement 
only makes sense when a (generalized) Boltzmann equation can be used. 
For example, if the optical response is determined by two distinct
charge carrier systems (two-component), so that the optical 
conductivity takes form:

\begin{equation}
\sigma(\omega)=\sigma^I(\omega)+\sigma^{II}(\omega),
\end{equation}

the interpretation of $M'(\omega)$ and $M''(\omega)$ as a scattering 
rate and a mass enhancement is meaningless, as can be seen 
from Eq.'s~7,8. This is the case if an interband transition is 
present in the same frequency region where there is an intraband 
response. We note however, that the form (12) can arise from a 
double-relaxation process (two different scattering mechanisms) as 
well.\cite{Allen_M} 

Since in the superconducting state $\sigma_1(\omega)$ is suppressed, the 
low-frequency optical conductivity is dominated by the imaginary term 
$\sigma_2(\omega)=\omega_{ps}^2/(4\pi\omega)$. In this case the 
low-frequency mass enhancement factor gives a ratio of the total plasma 
frequency, $\omega_p^2$, to the plasma frequency of the superconducting 
carriers, $\omega_{ps}^2$,: $1+\lambda(\omega)=\omega_p^2/\omega_{ps}^2$. 

\subsection{ Electron-boson scattering.}

Memory-function analysis has been most extensively developed for the 
case of electron-phonon scattering.\cite{Allen_M,Allen,Shulga} It can be 
shown in the limit of frequencies comparable to the Debye frequency 
and/or high enough temperature, the quasiparticle description breaks 
down.\cite{Shulga} Using more general many-body calculations Shulga {\it et 
al.} obtained the following expression for $1/\tau(\omega,T)$: 

\begin{eqnarray}
{1 \over \tau}(\omega,T)={\pi \over \omega} \int_0^{\infty} d\Omega 
\alpha_{tr}^2(\Omega)F(\Omega)[2{\omega}{\coth}({\Omega \over {2T}})-
(\omega+\Omega){\coth}({{\omega+\Omega} \over {2T}})+ \nonumber \\ +
(\omega-\Omega){\coth}({{\omega-\Omega} \over {2T}})] + 
{1 \over \tau_{imp}}.
\end{eqnarray}

Here $\alpha_{tr}^2(\Omega)F(\Omega)$ is a phonon density of states 
weighted by the amplitude for large-angle scattering on the Fermi 
surface and T is measured in frequency units. The last term in (13) 
represents impurity scattering. In the limit of zero temperature this 
reduces to Allen's result:\cite{Allen}

\begin{equation}
{1 \over \tau}(\omega)={{2\pi} \over \omega} \int_0^\omega d\Omega 
(\omega-\Omega) \alpha_{tr}^2(\Omega)F(\Omega) + {1 \over \tau_{imp}}.
\end{equation}

The dc scattering rate is obtained in the limit of $\omega=0$ in 
Eq.~13:

\begin{equation}
{1 \over \tau}(0,T) = \pi \int_0^{\infty} d\Omega 
\alpha^2_{tr}(\Omega)F(\Omega) {\Omega \over T} {\sinh}^{-2}({\Omega 
\over {2T}}) + {1 \over \tau_{imp}}.
\end{equation}

At temperatures much higher than the phonon spectrum upper-energy 
cut-off, $T{\gg}\Omega_{c}$, the above expression reduces to:

\begin{equation}
\lim_{T/\Omega_c\rightarrow \infty}{1 \over \tau}(0,T) = 4\pi T 
\int_0^{\infty} d\Omega {{\alpha^2_{tr}(\Omega)F(\Omega)}\over 
\Omega} + {1 \over \tau_{imp}},
\end{equation}
which is just the familiar result that the high-temperature 
electron-phonon contribution to a dc resistivity is linear in 
temperature.

In the limit of high $\omega$, $\omega{\gg}\Omega_{c}$,

\begin{equation}
\lim_{\omega/\Omega_c\rightarrow \infty}{1 \over \tau}(\omega,T) = 
2\pi \int_0^{\infty} d\Omega \alpha^2_{tr}(\Omega)F(\Omega) 
{\coth}({\Omega \over {2T}}) + {1 \over \tau_{imp}},
\end{equation}
which at high temperatures, $T{\gg}\Omega_c$, assumes the same value 
as the zero-frequency limit (16). Therefore, at very high 
temperatures the scattering rate becomes frequency-independent and 
the Eq.~6 reduces to the classical Drude expression (1).

We note that the effective scattering rate $1/\tau(\omega)$ is 
different from the quasiparticle attenuation $\gamma(\omega)$. For 
example, at zero temperature $\gamma(\omega)$ is given 
by:\cite{Allen,Millis}

\begin{equation}
{\gamma}(\omega)=-2Im\Sigma(\omega)={2\pi} \int_0^\omega d\Omega 
\alpha^2(\Omega)F(\Omega) + {1 \over \tau_{imp}}.
\end{equation}

Here $\alpha^2(\Omega)F(\Omega)$ is the isotropically weighted phonon 
density of states. One can see from Eq's~14,18 that at $T=0$ the 
effective scattering rate $1/\tau(\omega)$ is, if the difference 
between $\alpha_{tr}^2$ and $\alpha^2$ is neglected, an {\it average} 
of $\gamma(\omega)$ over frequencies from 0 to $\omega$ and therefore 
$\gamma(\omega)$ enters into the effective scattering rate in a way 
non-local in frequency.\cite{Shulga,Mori,Allen} As a consequence, 
$1/\tau(\omega)$ is actually equal to the quasiparticle attenuation 
$\gamma(\omega)$ only at $\omega=0$, where 
$1/\tau(0)=\gamma(0)=1/\tau_{imp}$. The two quantities also 
asymptotically approach each other in the limit of high frequencies, 
$\omega{\gg}\Omega_c$, where both $\gamma(\omega)$ and
$1/\tau(\omega)$ become frequency-independent. As the temperature is 
increased, the difference between $\gamma(\omega)$ and 
$1/\tau(\omega)$ is smeared out, and in the limit of $T{\gg}\Omega_c$ 
they are asymptotically equal. Generally, however, $\tau(\omega,T)$ 
deviates from the quasi-particle lifetime $\gamma^{-1}(\omega,T)$.

Eq's.~13,14, which have been derived for an electron-phonon 
scattering, are believed to be valid in the case of coupling of an 
electronic spectrum to any bosonic excitations.\cite{Webb,Shulga_1} 
In this case the Eliashberg function $\alpha_{tr}^2(\Omega)F(\Omega)$ 
in Eqs.~13,14 is replaced by the corresponding, suitably weighted, 
bosonic spectral density ${\cal A}_{tr}(\omega)$. To give a flavor of 
the results expected on the basis of Eq's~13,14 we will perform 
calculations for several model shapes of ${\cal A}_{tr}(\omega)$: a 
$\delta$-peak, a "square"-like spectrum and ${\cal 
A}_{tr}(\omega)=\Gamma\omega/(\Gamma^2+\omega^2)$. The last spectrum 
is believed to be appropriate for scattering of electrons on spin 
fluctuations.\cite{millis90} 

\begin{figure}[htpb]
\leavevmode
\epsfxsize=0.50\columnwidth
\centerline{\epsffile{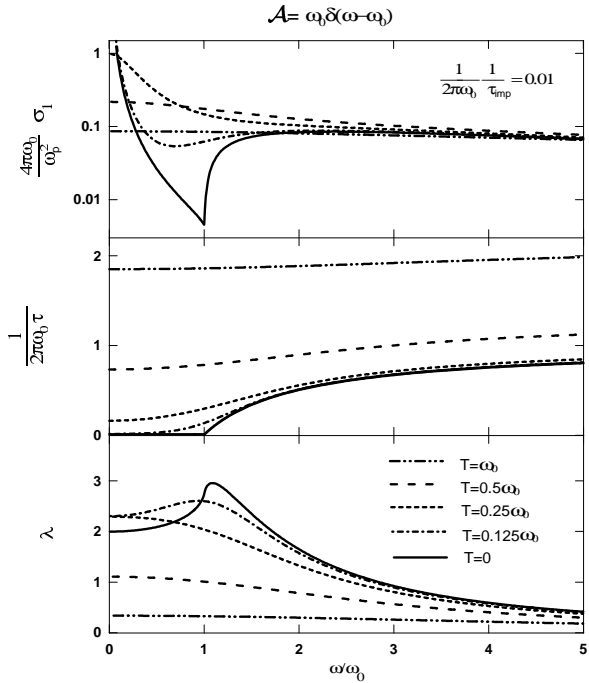}
 }
 \caption{Electron-boson model calculations with boson spectral 
density ${\cal A}_{tr}(\omega)=\omega_0\delta(\omega-\omega_0)$. Top panel 
is the optical conductivity, middle panel is the scattering rate and 
bottom panel is the mass renormalization. The coupling constant is 
equal to 1.}
\label{theory1}
\end{figure}

In the case of ${\cal 
A}_{tr}(\omega)=\omega_0\delta(\omega-\omega_0)$ the integration of 
Eq.~13 can easily be done. The effective mass enhancement 
$\lambda(\omega)$ can be calculated using the Kramers-Kronig relation 
(9). As soon as both $1/\tau(\omega,T)$ and $\lambda(\omega,T)$ are 
known, the optical conductivity can be calculated using Eq.~3. The 
impurity scattering has been set to 
$1/(2{\pi}\omega_0\tau_{imp})=0.01$. The results obtained are 
presented in Fig.~\ref{theory1} at five different temperatures: $T=0, 
0.125\omega_0, 0.25\omega_0, 0.5\omega_0, \omega_0$.

\begin{figure}[htpb]
\leavevmode
\epsfxsize=0.50\columnwidth
\centerline{\epsffile{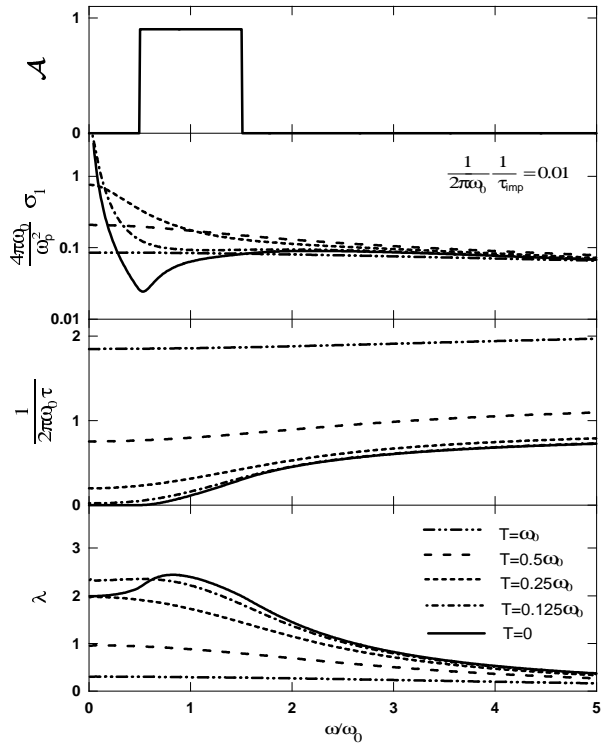}
 }
 \caption{Electron-boson model calculations. Top panel shows the 
bosonic spectral density, here a "square" spectrum, next panel is the 
optical conductivity, next 
panel is the scattering rate and the bottom panel is the mass renormalization.
The coupling constant is equal to 1.}
\label{theory2}
\end{figure}

\begin{figure}[htpb]
\leavevmode
\epsfxsize=0.50\columnwidth
\centerline{\epsffile{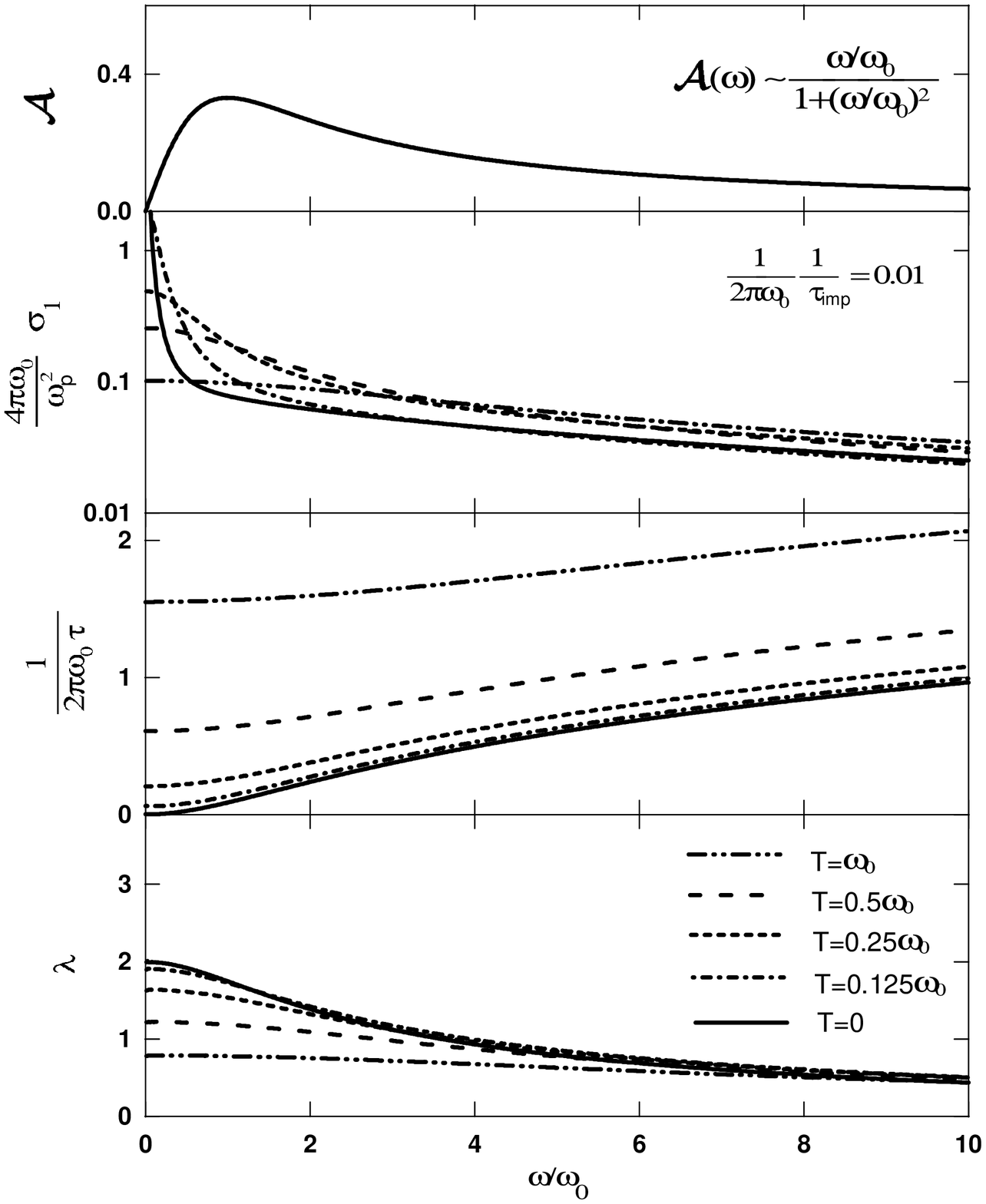}
 }
 \caption{Electron-boson model calculations. Top panel shows the 
bosonic spectral density, next panel is 
the optical conductivity, next panel 
is the scattering rate and the bottom panel is the mass renormalization. 
The coupling constant is equal to 1.}
\label{theory3}
\end{figure}

For the two other choices of ${\cal A}_{tr}(\omega)$, the integration 
of Eq.~13 was done numerically. We then used the Kramers-Kronig 
relation to obtain $\lambda(\omega,T)$. The same impurity 
scattering rate as in the case of $\delta$-function was used to calculate 
$\sigma_1(\omega)$. The results are presented in Fig.~\ref{theory2} and 
Fig.~\ref{theory3} at different temperatures, measured in units of 
the characteristic frequency of bosonic spectrum ${\cal 
A}_{tr}(\omega)$.

As was discussed above, if the ${\cal A}_{tr}(\omega)$ spectrum has a 
high-energy cut-off, $1/\tau(\omega,T)$ saturates at frequencies that 
are much higher than the cut-off. The value of $1/\tau(\omega,T)$ in 
the saturation regime is strongly temperature-dependent, and linear 
in T at high enough temperatures according to (17). However, if 
there is no cut-off in ${\cal A}_{tr}(\omega)$, as in the case of 
magnetic spectrum in Fig.~\ref{theory3}, there is no high-frequency 
saturation of $1/\tau(\omega,T)$, rather it continues to increase. 
The temperature dependence of the absolute value of the scattering 
rate is, however, still strong. 

In Fig.~\ref{theory1}, the effective mass enhancement 
$\lambda(\omega,T)$ is larger at low frequencies and decreases to 
zero at high frequencies. This has a simple physical interpretation 
that at high frequencies the boson "cloud" is not capable of 
following the charge carriers. The sharp increase in the 
low-temperature $\lambda$ around the frequency of bosonic excitation 
$\omega_0$ corresponds to the onset of boson-emission process, since 
only carriers with energy greater than $\hbar\omega_0$ can emit a 
boson. A similar onset can be seen in Fig.~\ref{theory2}. In the case 
of the magnetic ${\cal A}_{tr}(\omega)$, this feature is not observed 
since bosons can be emitted by a carrier with arbitrarily small 
energy. At high temperatures $\lambda$ asymptotically approaches 
zero, in agreement with the frequency-independent scattering rate 
$1/\tau$. 

The low-temperature conductivity in Fig.~\ref{theory1} shows a 
pronounced absorption band, called a Holstein band, with a sharp 
onset at $\omega_0$. The band corresponds to an additional absorption 
channel associated with boson emission processes. Similar absorption 
onset can be seen in Fig.~\ref{theory2} but not in 
Fig.~\ref{theory3}. The reason for this, as in the case of $\lambda$, 
is the large boson spectral density at all non-zero frequencies for 
the magnetic ${\cal A}_{tr}(\omega)$. As the temperature is 
increased, all sharp features in $\sigma_1$ are smeared out and at 
very high temperatures the conductivity can be described by a single 
Lorentzian of Eq.~1.

\section{EXPERIMENTAL RESULTS}

This section is divided into three subsections: underdoped, optimally 
doped and overdoped. In each of the subsections we first present the 
raw experimental results in the form of absorption 
$A(\omega)=1-R(\omega)$ for a selected material at many different 
temperatures. Before we proceed to the memory function analysis, we 
will also present the results for the same material in terms of more 
commonly used real optical conductivity $\sigma_1(\omega)$. However, 
we will focus the analysis on the real and imaginary parts of the 
memory function $M(\omega)=M'(\omega) + i M''(\omega)$, that will be 
presented for several materials on a second diagram in each 
subsection. While for a selected material in each subsection we will 
show many different temperatures, to simplify the diagrams for 
others, only three temperatures will be shown: T=300~K, 
$T{\simeq}T_c$ and the lowest (superconducting) temperature.

We note here again that we are fully aware that in most real 
situations, and especially in HTSC, the real and imaginary parts of 
$M(\omega)$ are not solely determined by the scattering effects and 
the corresponding enhancement of an effective mass. Nevertheless, 
mostly for historical reasons, we will refer to the effective 
scattering rate and to the effective mass defined as 
$1/\tau(\omega)=M'(\omega)$ and 
$m^*=1+\lambda(\omega)=1-M''(\omega)/\omega$ respectively. Keeping 
this in mind, we will now present the experimental results and 
indicate the common trends, leaving the interpretation for the 
discussion section. Since we will mainly be interested in the 
evolution of the optical response in the pseudogap energy region we 
will present the experimental data up to 2000~cm$^{-1}$ only.

\subsection{Underdoped cuprates.}

\begin{figure}[htpb]
\leavevmode
\epsfxsize=0.75\columnwidth
\centerline{\epsffile{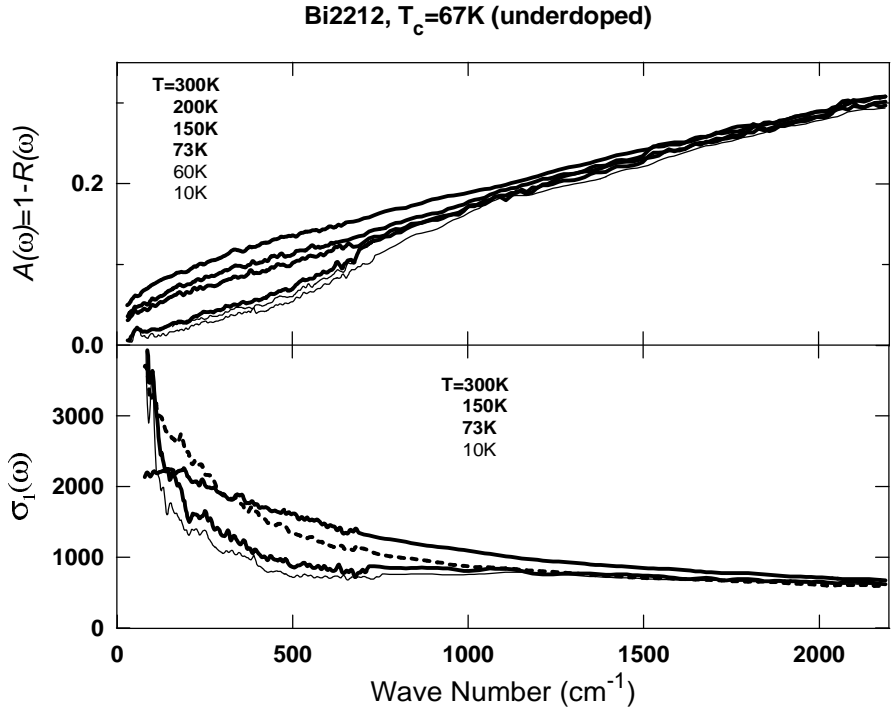}
 }
 \caption{The absorption $A=1-R$, top 
panel, and the optical conductivity $\sigma_1(\omega)$ for 
underdoped B2212 ($T_c=67$~K). The absorption rises linearly at high 
temperatures but develops a depression below 800 cm$^{-1}$ due to the 
formation of the pseudogap. In the optical conductivity the pseudogap 
shows up as a narrowing of the coherent Drude peak at low frequency.} 
\label{under-abs}
\end{figure}

A typical plot of the temperature dependence of raw absorption data 
$A(\omega,T)$ for an underdoped HTSC is shown in 
Fig.~\ref{under-abs}, this particular example being underdoped Bi2212 
material with $T_c=67$~K. In the temperature range 300-150~K the 
absolute value of the low-frequency absorption decreases smoothly 
with decreasing temperature without any sharp features. 
However, at a temperature $T<T^*{\simeq}150$~K, the absorption 
below 600-700~cm$^{-1}$ starts to decrease faster than at higher 
frequencies, developing a threshold structure which is characteristic 
for an underdoped HTSC in the pseudogap state.

The corresponding changes in the real part of optical conductivity 
$\sigma_1(\omega)$ are also shown in Fig.~\ref{under-abs}
at selected temperatures. The in-plane response of all 
samples is metallic, {\it i.e.} the absolute value of 
$\sigma_{1}(\omega)$ decreases from the dc value with increasing 
$\omega$. However, while the $\sigma_1(\omega)$ spectra are quite 
broad at temperatures above $T^*$, the rapid decrease of the 
low-frequency absorption below $T^*$ results in an abrupt narrowing 
of the low-frequency conductivity with substantial spectral weight 
being transferred towards zero frequency. As temperature is reduced 
below $T_c$, no dramatic changes are observed in the optical response 
of underdoped cuprates: the only change is just a {\it continued} 
narrowing of the intense low-frequency peak, that has already 
initiated in the normal state.

\begin{figure}[htpb]
\leavevmode
\epsfxsize=0.94\columnwidth
\centerline{\epsffile{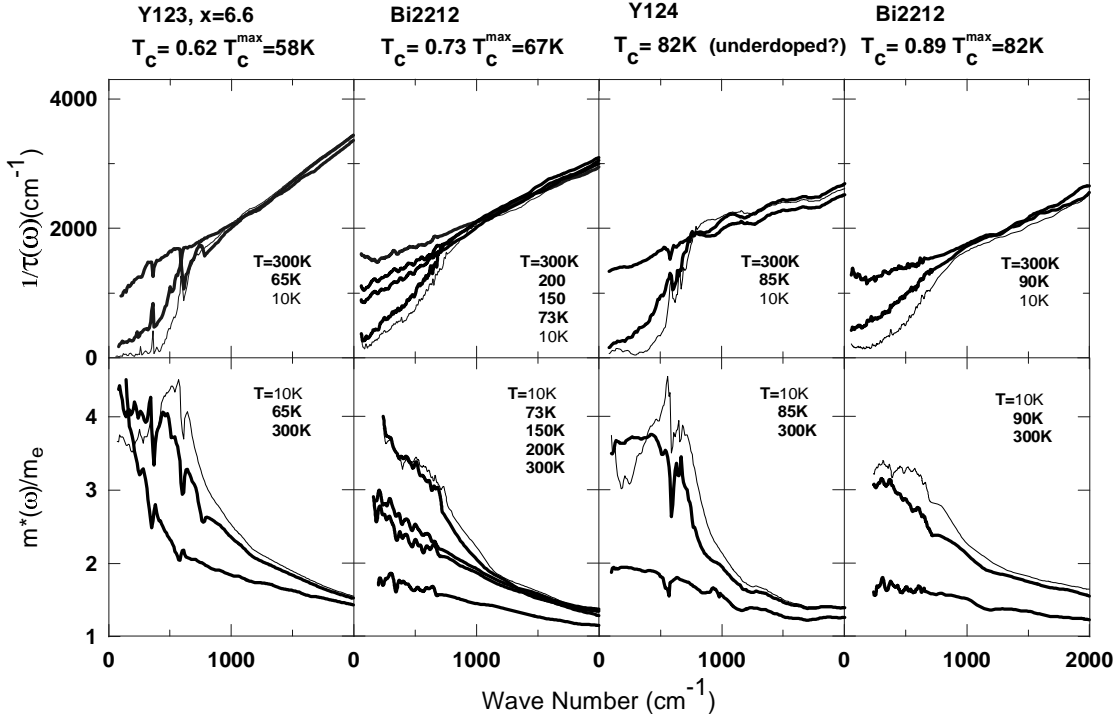}
 }
 \caption{The frequency dependent scattering rate, top row, and 
the mass renormalization for a series of underdoped cuprate 
superconductors, bottom row. The scattering rate curves are 
essentially temperature independent above 1000 cm$^{-1}$ but develop 
a depression at low temperature and low frequencies. The effective 
mass is enhanced at low temperature and low frequencies.}
\label{under-tau}
\end{figure}

The scattering rate $1/\tau$ and the effective mass 
$m^*/m_e=1+\lambda$ for the Bi2212 crystal with $T_c=67$~K, 
calculated from the optical conductivity using the formulae described 
in section III, are shown in Fig.~\ref{under-tau}. 
We have used a plasma frequency value of $\omega_p=14300$~cm$^{-1}$, 
obtained by using the conductivity sum-rule 
analysis\cite{timusk1,puchkov96a} with integration of 
$\sigma_1(\omega)$ over all frequencies up to 1~eV. We note that the 
value of $\omega_p$ obtained this way is somewhat ambiguous since 
there is no clear separation between the frequency regions of the 
free- and bound-carrier optical responses. However, a particular 
choice of $\omega_p$ only multiplies $1/\tau(\omega)$ and 
$m^*(\omega)$ by a constant. Since in this paper we are mostly interested in the 
{\it frequency dependence} of these quantities, the exact value of 
$\omega_p$ is not of primary importance. To keep the absolute 
values consistent throughout the paper, in Bi2212 and Tl2201 series 
we will use plasma frequency values obtained by integrating the real 
part of optical conductivity up to 1~eV, which seems to be an 
energy below which the conductivity is substantially changed by 
doping.\cite{orenstein90,puchkov96a} In YBCO series we will use an 
energy of 1.5~eV as an upper integration limit since the reflectivity 
plasma minimum is higher for these materials.\cite{timusk1} 

The scattering rate $1/\tau(\omega)$ of underdoped Bi2212 with 
$T_c=67$~K is linear at frequencies from 800~cm$^{-1}$ to at least 
3000~cm$^{-1}$ at all temperatures. While at room temperature the 
low-frequency $1/\tau(\omega)$ deviates upwards from the 
high-frequency linear law, at $T=200K$ the spectrum is linear over 
the whole frequency range from 100-3000~cm$^{-1}$. However, as 
temperature is reduced below $T^*$, the scattering rate is suppressed more 
rapidly at low frequencies ($\omega<700$ cm$^{-1}$) while it remains nearly 
unaffected at higher energies. A result of this development is an 
appearance of a distinct threshold in the $1/\tau(\omega)$ spectra. 
Another interesting phenomenon, that we will return to later, is the 
remarkable temperature-independence of the high-frequency 
$1/\tau(\omega)$.

Similar to other quantities, the effective mass $m^*(\omega)$ 
displays a rapid change at frequencies below 
700~cm$^{-1}$ as temperature is reduced below $T^*$. We note that 
the narrowing of the low-frequency optical conductivity is a result 
of {\it both} a decrease of $1/\tau(\omega)$ and an increase of 
$m^*(\omega)$ since heavy carriers are more difficult to scatter. The 
width of a conductivity peak is determined by a renormalized 
scattering rate $1/\tau^*(\omega)=m_e/(\tau(\omega)m^*(\omega))$. At 
low temperatures the effective mass saturates at $m^*(0){\simeq}3-4$. 

The experimental results obtained for several other cuprate materials 
at different doping levels in the underdoped state are qualitatively 
similar. In the rest of Fig.~\ref{under-tau}. we show the effective 
scattering rate and the carrier mass obtained for Y123 with oxygen 
content $x=6.6$ and $T_c=58$~K, naturally underdoped Y124 with 
$T_c=80$~K, and slightly underdoped Bi2212 with $T_c=82$~K. The 
in-plane plasma frequency $\omega_p$, related to the conductivity by 
$\omega_p^2/8 = \int_0^\infty \sigma_1(\omega) d\omega$, scales with 
$T_c$ in accordance with earlier 
measurements.\cite{orenstein90,puchkov96a} Integration of the 
conductivity up to 1.5~eV yields the following values of the plasma 
frequency: 15000~cm$^{-1}$ in YBa$_2$Cu$_3$O$_{6.6}$, 16000~cm$^{-1}$ 
in YBa$_2$Cu$_4$O$_8$ and 15600~cm$^{-1}$ in Bi2212. For clarity, 
only three temperatures are shown for each material: room 
temperature, just above $T_c$ and well below $T_c$. 

All of the samples show the same characteristic suppression of the 
amplitude of the scattering rate at $T<T^*$, which seems to 
increase as doping level decreases. Despite the differences in 
the values of $T^*$ in the different samples, the energy scale 
associated with the suppression of $1/\tau(\omega)$, 
does not change significantly with doping. 
In particular, a deviation from the linear behavior in all 
studied samples occurs at the same frequency $\omega<700$~cm$^{-1}$. 
As the doping level is increased towards the optimal, the normal 
state depression of $1/\tau(\omega)$ becomes progressively shallower, 
while in the superconducting state the depression remains almost 
unchanged. The net effect is that the difference between the 
low-temperature normal-state and the superconducting-state 
$1/\tau(\omega)$ becomes more prominent as doping level approaches 
the optimal. At the same time, qualitatively, the shape of the 
normal-state $1/\tau(\omega)$ at $T<T^*$ remains similar to that in 
the superconducting state. With an exception of Y124 sample, the 
high-frequency $1/\tau(\omega)$ is linear up to at least 
3000~cm$^{-1}$ (2000~cm$^{-1}$ for Y124) and for all samples it is 
nearly temperature-independent. The low-temperature effective mass 
$m^*(\omega)$ becomes enhanced at low frequencies when temperature is 
reduced below $T^*$. In all samples $m^*(\omega)$ saturates at about 
the same value of ${\approx}3-4$. 

To summarize, the optical response of underdoped cuprates is 
characterized by the following generic features: (i) the scattering 
rate is nearly linear with $\omega$ at $T>T^*$; (ii) At $T<T^*$ 
(the pseudogap state) the low-frequency scattering rate is suppressed 
corresponding to the rapid narrowing of the Drude-like feature in the 
conductivity spectra. The energy scale associated with the changes of 
$1/\tau(\omega)$ spectra was found to be the same in all samples. 
The magnitude of the depression weakens as doping is increased 
towards the optimal level. (iii) The high-frequency $1/\tau(\omega)$ 
remains effectively temperature-independent and linear from 
700~cm$^{-1}$ up to at least 3000~cm$^{-1}$ in most underdoped HTSC 
samples.

\subsection{Optimally doped and lightly overdoped cuprates.}

\begin{figure}[htpb]
\leavevmode
\epsfxsize=0.70\columnwidth
\centerline{\epsffile{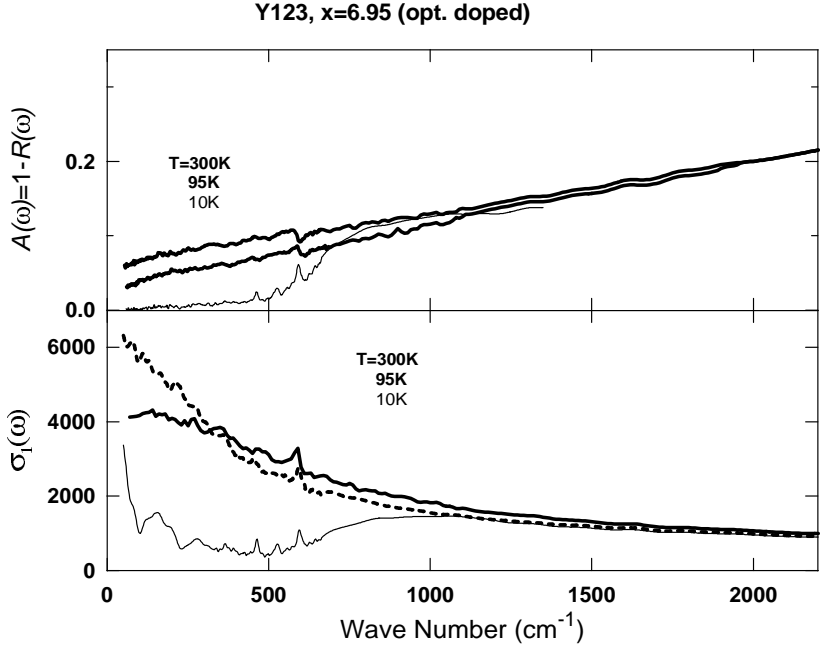}
 }
 \caption{The absorption $A=1-R$ and the optical conductivity for 
optimally doped Y123 with $x=6.95$ ($T_c=93.5$~K). 
A depression of A is seen below 800 cm$^{-1}$ but only below the 
superconducting transition temperature $T_c$. The same is true for 
the characteristic narrowing of the optical conductivity.}
\label{opt-abs}
\end{figure}

\begin{figure}[htpb]
\leavevmode
\epsfxsize=0.7\columnwidth
\centerline{\epsffile{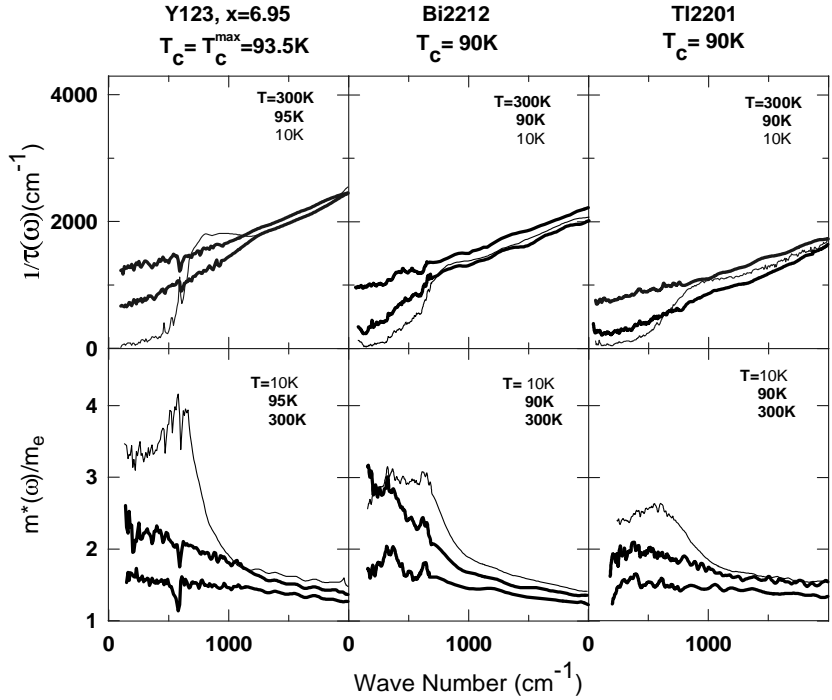}}
\caption{The scattering rate and the effective mass for samples
close to optimal doping. The scattering rate now has a degree of
temperature dependence at low frequencies. In the superconducting
state the scattering rate is depressed at low frequencies.
}
\label{opt-tau}
\end{figure}

A similar threshold structure in the raw absorption spectra is 
observed in the optimally doped crystals as well. As an example, in 
Fig.~\ref{opt-abs} we show absorption and conductivity 
data obtained from optimally doped Y123 
material. The important difference from the underdoped cuprates is 
that now a threshold in $A(\omega)$ develops only at temperatures 
below $T_c$. The corresponding $1/\tau(\omega)$ 
and $m^*(\omega)$ spectra are plotted in Fig.~\ref{opt-tau}. 
We have used a plasma frequency 
${\omega}_p=18000$~cm$^{-1}$, obtained from the sum-rule analysis with 
integration up to 1.5~eV. All of the optical constants show the same 
characteristic features as in underdoped cuprates but the onset 
temperature is determined now by $T_c$. A remarkable feature of the 
optimally doped samples is the similarity between the behavior of the 
superconducting-state optical response obtained in these crystals 
with the data obtained in the underdoped materials at 
$T_c<T<T^*$. This would be consistent with the notion 
that the $T_c$ and $T^*$ boundaries in Fig.~\ref{phase-dia1} cross 
around the optimal doping. As a result, the difference between the 
normal-state and the superconducting state spectra becomes dramatic 
in optimally doped samples. 

In the normal state, as the temperature is reduced from 300~K down to 
$T{\simeq}T_c$, both the scattering rate and the renormalized 
effective mass, in optimally doped samples, show relatively minor 
changes. These changes are mainly restricted to the decrease of the 
absolute value of $1/\tau(\omega)$ in the low frequency parts of the 
spectra. However, in contrast to the underdoped materials, the 
normal-state scattering rate in Y123 does not reveal any sharp 
changes in the frequency dependence as temperature is reduced.

In the rest of Fig.~\ref{opt-tau} we show data obtained on Bi2212 
with $T_c=90$~K and Tl2201 with $T_c=90$~K. We should note that 
although we assigned the material Tl2201 to this section, the peak in 
$T_c$ as a function of doping has not being observed for Tl2201 and 
some data suggest that this material may be somewhat 
overdoped.\cite{puchkov96a} The plasma frequency used for Bi2212 was 
$\omega_p=16000$~cm$^{-1}$ and for Tl2201 $\omega_p=15300$~cm$^{-1}$. 
The normal-state spectra of $1/\tau(\omega)$ are featureless. In case 
of Tl2201 the threshold structure appears only at $T<T_c$ but in 
Bi2212 a weak structure can still be seen at T=90~K. In fact, it 
persists even in the lightly overdoped samples. Thus it is possible 
that the pseudogap state in Fig.~\ref{phase-dia1} can somewhat 
penetrate into the overdoped regime. Qualitatively, the 
depression in $1/\tau(\omega)$ at $T<T_c$ in the optimally doped 
cuprates is very similar to what is observed in the $1/\tau(\omega)$ 
spectra in the pseudogap state of the underdoped cuprates. However in
contrast with the underdoped materials, the temperature dependence of 
the scattering rate now seems to extend over a broader frequency 
range. In particular, in the Bi2212 and Tl2201 samples the 
$1/\tau(\omega)$ spectra reveal some shift in the high-frequency part 
(above the 700 cm$^{-1}$ threshold) whereas in the underdoped 
materials no temperature dependence was observed at these 
frequencies. 

Another weak feature that seems to be common for both the optimally 
doped Y123 and $T_c=90$~K Tl2201 is an "overshoot" of the 
superconducting-state $1/\tau(\omega)$ above the spectrum of 
$1/\tau(\omega)$ for $T{\simeq}T_c$.

In summary, the response of the optimally doped high-$T_c$ cuprates 
demonstrates the following features: (i) A threshold feature in 
$1/\tau(\omega)$ spectra at $T>T_c$ is either strongly suppressed or disappears 
completely when doping level approaches optimal; (ii) The 
high-frequency $1/\tau(\omega)$ remains linear but may acquire a 
weak temperature dependence in lightly overdoped cuprates.

\subsection{ Overdoped cuprates.}

\begin{figure}[htpb]
\leavevmode
\epsfxsize=0.7\columnwidth
\centerline{\epsffile{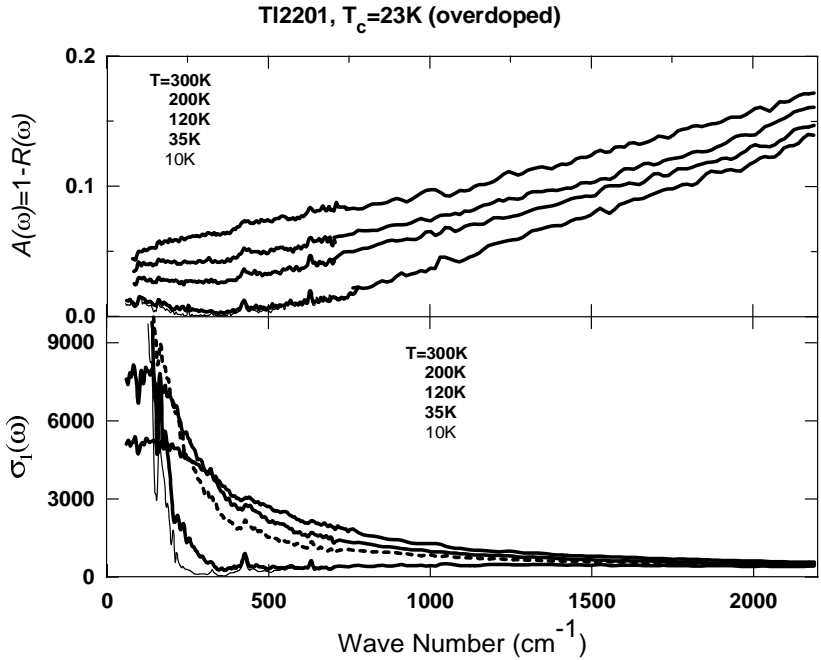}
 }
\caption{The absorption of strongly overdoped Tl2201 ($T_c=23$~K), 
top panel and the optical conductivity, lower panel. The absorption 
is strongly temperature dependent but no threshold develops at low 
temperatures. The optical conductivity becomes narrower as 
temperature decreases but does not show any sharp changes.}
\label{over-abs}
\end{figure}

Since the strongly overdoped regime is not accessible in the 
Bi2212 or in the YBCO materials, we have chosen Tl2201, (Bi/Pb)2212 
and slightly overdoped Bi2212 in order to study this doping regime. 
In Fig.~\ref{over-abs} we show the data for a strongly overdoped 
high-$T_c$ superconductor (Tl2201 with $T_c=23~$K). The raw 
absorption spectra are qualitatively different from those obtained in 
optimally doped or underdoped regimes. The $A(\omega)$ is 
temperature-dependent over a much broader frequency range. The 
spectra shift down uniformly as temperature decreases but no sharp 
features develop. Unfortunately, in this crystal absorption is 
already very small in the normal state at $T=35$~K. It is difficult
to determine the exact shape of $A(\omega)$ in the superconducting 
state. Thus it remains unclear if the absorption spectra of this 
crystal shows the same threshold structure as the less heavily 
doped materials. 

\begin{figure}[htpb]
\leavevmode
\epsfxsize=0.7\columnwidth
\centerline{\epsffile{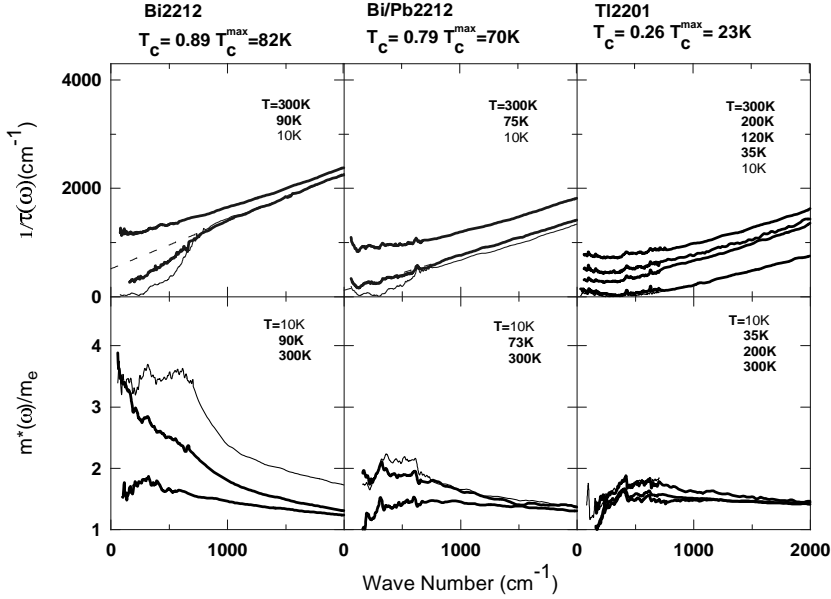}
 }
 \caption{In overdoped samples the scattering rate, top panels, shows 
an increasingly strong temperature dependence. As a part of the 
high-frequency scattering disappears at low temperatures the 
low-frequency depression of $1/\tau(\omega)$ and the effective mass 
enhancement decreases in magnitude, even in the superconducting 
state.}
\label{over-tau}
\end{figure}

The $\sigma_1(\omega)$ spectra for the strongly overdoped Tl2201 are 
shown in the bottom panel of Fig.~\ref{over-abs} while $1/\tau(\omega)$ and 
$m^*(\omega)$ spectra are shown in Fig.~\ref{over-tau}. The 
plasma frequency is $\omega_p=15100$~cm$^{-1}$. Consistent with the 
behavior of the absorption spectra there is no sharp change in the 
frequency dependence in any of these response functions as 
the temperature is decreased in the normal state. Instead, the 
$1/\tau(\omega)$ spectra scale downwards almost parallel to each 
other. This is in a sharp contrast with the
$1/\tau(\omega)$ behavior in the underdoped regime, where 
the scattering rate was found to be temperature independent above 
1000~cm$^{-1}$. We also note that the frequency dependence of 
$1/\tau(\omega)$ for this strongly overdoped material may become 
superlinear, flattening out at low frequencies. The effective mass 
$m^*(\omega)$ does not show any pronounced temperature dependence and 
remains largely flat in the whole frequency region shown. To show 
the continuity in the evolution of the optical response of the 
cuprates from under- and optimally doped to the strongly overdoped 
case we plot $1/\tau(\omega)$ and $m^*(\omega)$ spectra for
other overdoped samples in the rest of Fig.~\ref{over-tau}. These 
include Bi2212 ($T_c=82$~K) and (Bi/Pb)2212 ($T_c=70$~K) annealed in 
oxygen ($\omega_p=15600$~cm$^{-1}$ for Bi2212 and 16500~cm$^{-1}$ for 
(Bi/Pb)2212). As we have noted in the previous section, the 
$1/\tau(\omega)$ spectrum for the slightly overdoped Bi2212 still 
shows a weak normal-state pseudogap feature at T=90~K, defined as a 
downwards deviation from the linear high-frequency behavior. However, 
(Bi/Pb)2212 shows no sign of a threshold formation above $T_c$. While 
the scattering rate remains close to linear in $\omega$ at high 
frequencies, it seems to gradually pick up a temperature dependence 
as the doping level is increased from the optimal to overdoped. Also, 
the absolute value of the scattering rate is gradually suppressed 
with increased doping. 

In the superconducting state the threshold structure seems to weaken as 
doping is increased towards strong overdoping. Correspondingly, the 
superconducting-state mass enhancement also becomes weaker. 
Unfortunately, as in case of absorption, we can not unambiguously 
determine the exact nature of changes that occur below $T_c$ in 
either $1/\tau(\omega)$ or $m^*(\omega)$ for the Tl2201 sample with 
$T_c=90$~K.

In summary, as doping level is increased above optimal to overdoped 
and strongly overdoped levels: (i) No threshold is observed in 
$1/\tau(\omega)$ at $T>T_c$. (ii) The scattering rate 
$1/\tau(\omega)$ acquires temperature dependence over a much broader 
frequency range than in underdoped cuprates. (iii) The frequency 
dependence of $1/\tau(\omega)$ may become superlinear in the
strongly overdoped cuprates.

\subsection{The effect of zinc doping}

\begin{figure}[htpb]
\leavevmode
\epsfxsize=0.80\columnwidth
\centerline{\epsffile{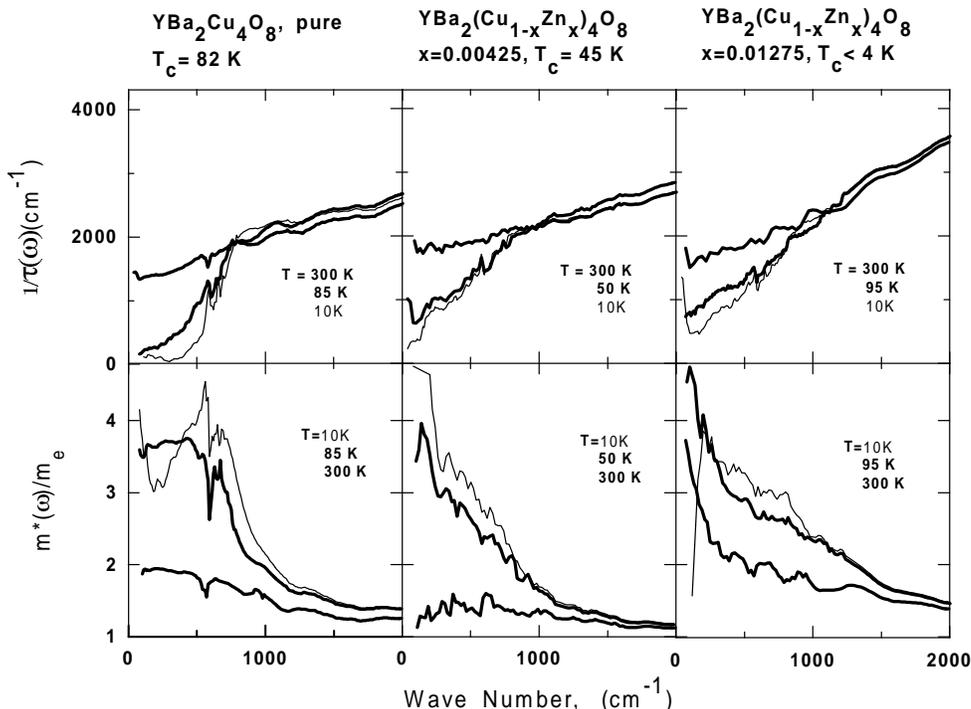}
 }
 \caption{Effect of zinc doping on the $ab$-plane scattering rate for 
Y124. The left panels show a pure sample while in the middle and 
right panels results for Y124 with 0.425$\%$ and 1.275$\%$ of Zn, 
respectively, are shown. Note the almost complete elimination of the 
threshold structure at ${\sim}700$~cm$^{-1}$. }
\label{Zn}
\end{figure}

In Fig.~\ref{Zn} we show the spectra of the scattering rate and the 
effective mass for a pure crystal of Y124 and for two samples 
containing 0.425~$\%$ and 1.275~$\%$ of Zn. As the result of Zn 
substitution, $T_c$ is suppressed from 82 K in pure crystal down to 
45 K in the material with 0.425~$\%$ of Zn. In the sample with 
1.275~$\%$ Zn, superconductivity is not observed above 4 K. It is 
believed that Zn substitutes Cu atoms primarily in the CuO$_2$ 
planes. 

In the crystals with Zn substitution the scattering rate is enhanced 
over the whole energy scale. The frequency dependence of 
$1/\tau(\omega)$ is modified as well. In particular, the threshold 
structure at $\omega\simeq 600$ cm$^{-1}$ weakens with increasing Zn 
content. The behavior of the non-superconducting crystal is in fact 
very similar to what is observed in the optimally doped samples at 
$T>T^*$. The principle difference for the Zn-doped underdoped 
crystals and optimally doped materials is that the threshold 
structure in $1/\tau(\omega)$ does not appear even below $T_c$.

At the low-frequencies we observe an upturn in 
the $1/\tau(\omega)$ spectra. A similar upturn is also observed in 
the disordered (Bi/Pb)2212 system and in Tl2201.\cite{puchkov95} It 
is likely that this behavior of the scattering rate could be 
attributed to incipient localization in the CuO$_2$ planes initiated 
by impurities. The upturn becomes stronger as the temperature 
decreases. We note that the dc properties of these materials, and in 
particular the temperature dependence of the dc resistivity, are 
determined by $1/\tau(\omega)$ at $\omega\rightarrow 0$. Thus an 
obvious consequence of the low-frequency upturn would be to reduce 
the slope of the $d\rho/dT$ dependencies and to create a residual 
resistivity. 

\section{Discussion} 

\subsection{General trends in $1/\tau(\omega)$ data.}

With regards to the underdoped cuprates, two distinct features in the 
$1/\tau(\omega)$ spectra deserve mentioning. First, it must be 
recognized that $1/\tau(\omega)$ is linear and almost temperature 
independent at high frequencies. Second, a threshold structure 
develops at low frequencies and temperatures below $T^*$.
When the doping 
reaches the optimal level the threshold structure in the ab-plane 
scattering rate shows up only in the superconducting state. This is 
accord with the phase diagram where the two curves, the pseudogap 
boundary and the superconducting transition temperature $T_c$, cross 
at optimal doping ($T^*{\leq}T_c$). In the overdoped cuprates the 
threshold structure appears only below $T_c$ and seems to become 
weaker even in the superconducting state as doping progresses. 
Unfortunately, the limitations of our experiment do not allow us to 
say with certainty if the structure persists in the strongly 
overdoped materials. The important difference between 
$1/\tau(\omega)$ for overdoped and underdoped materials is a strong
temperature dependence of the high-frequency part of 
$1/\tau(\omega)$ in the overdoped case.

The common feature in all spectra is the linear dependence of the 
high-frequency scattering rate. The linear frequency dependence has 
been seen previously in the scattering rate of the $a$-axis Y123 both 
in the optimally doped and underdoped 
spectra.\cite{schlesinger90,rotter91} In Table.~1 we present the 
slopes and zero-frequency intercepts of the high-frequency part of 
scattering rate obtained by fitting it to a straight line 
$1/\tau(\omega)=\alpha\omega+\beta$. The results are presented at two 
temperatures: T=300~K and at the lowest normal temperature (in 
parentheses). 

We note here that the coefficients determined directly from 
$1/\tau(\omega)$ may be ambiguous since they involve the plasma 
frequencies that were obtained by integrating the real part of 
conductivity up to a somewhat arbitrary chosen frequency. However, 
the same cut-off integration frequency was used for each of the 
series at all doping levels (1.5~eV for YBCO and 1~eV for Bi2212 and 
Tl2201). While the absolute value of $\omega_p$ obtained in this 
manner may still be ambiguous, the {\it changes} in $\omega_p$ with 
doping reflect changes in the carrier density for each of
material series.\cite{puchkov96a} For these reasons the materials 
presented in Table.~1 are grouped by series. Another way to get 
around the problem of the unknown plasma frequency is to divide
scattering rate by $\omega_p^2$: $4\pi/(\omega_p^2\tau)$. This 
quantity may be called "optical resistivity", or $\rho_{opt}$, since 
it has the same functional form as a dc resistivity in a simple Drude 
model.\cite{Ashcroft} Since it is directly obtained from the 
measured complex optical conductivity: 
$\rho_{opt}(\omega)=Re(1/\sigma(\omega))$, it may be useful to 
examine variations of the slope and zero-frequency intercept of 
$\rho_{opt}(\omega)$ instead of $1/\tau(\omega)$. The corresponding 
results are listed in the last two columns of Table.~1.

\vspace{0.1in}
\begin{table}[htpb] 
\caption{The slopes and zero-frequency intercepts of the 
high-frequency linear part of $1/\tau(\omega)=\alpha\omega+\beta$, 
third and fourth columns. The linear coefficients normalized to the 
plasma frequency, fifth and sixth columns. The fit was performed over 
a frequency range from 900-3000~cm$^{-1}$. The Y124 material is not 
shown since the high-frequency scattering rate significantly deviates 
from linear above 2000~cm$^{-1}$.}

\begin{tabular}{|c|c||c|c||c|c|} \hline

Material & $T_c$ & $\alpha$ & $\beta$ (cm$^{-1}$) & 
$4\pi\alpha/\omega^2_p$ ($\mu\Omega$cm$^2$) & $4\pi\beta/\omega^2_p$ ($\mu\Omega$cm)\\
\hline 
Y123 (u.d) & 58~K & 1.26 (1.45) & 790 (560) & 0.34 (0.39) & 210 (149) \\
Y123 (opt.d) & 93.5~K & 0.79 (0.93) & 890 (590) & 0.15 (0.17) & 165 (108) \\
\hline 
Bi2212 (u.d.) & 67~K & 0.84 (0.91) & 1280 (1200) & 0.25 (0.27) & 377 (352) \\
Bi2212 (u.d) & 82~K & 0.76 (0.95) & 990 (750) & 0.19 (0.23) & 243 (185) \\
Bi2212 (opt.d.) & 90~K & 0.71 (0.72) & 850 (650) & 0.17 (0.17) & 200 (150) \\
Bi2212 (o.d.) & 82~K & 0.73 (0.77) & 890 (550) & 0.18 (0.19) & 219 (135) \\
(Bi/Pb)2212 (o.d.) & 70~K & 0.63 (0.65) & 551 (118) & 0.13 (0.14) & 117 (25) \\
\hline 
Tl2212 (o.d.?) & 90~K & 0.64 (0.75) & 473 (90) & 0.16 (0.19) & 121 (23) \\
Tl2212 (o.d.) & 23~K & 0.63 (0.54) & 337 (-318) & 0.17 (0.14) & 89 (-84) \\
\hline 
\end{tabular} 
\label{table1} 
\end{table}

The result of both approaches is that both $\alpha$ and $\beta$ seem 
to decrease with doping for all of the series. However, while the 
decrease in the slope is insignificant (and may even be inside our 
error bar estimated to be about 20\%), the drop in the 
intercept, especially its low-temperature value, is dramatic. We 
also note the large difference between the room-temperature and 
low-temperature (numbers in parenthesis) intercept values in the 
overdoped cuprates, which is a result of the intense temperature 
dependence of the high-frequency part of $1/\tau(\omega)$. 
The low-temperature intercept even becomes negative for strongly 
overdoped Tl2201. 

The low intercept values in overdoped cuprates suggest that the 
temperature dependence and the low-frequency threshold in 
$1/\tau(\omega)$ are closely related. The intense 
temperature-induced suppression of $1/\tau(\omega)$ over a large 
frequency range makes the high-frequency background at $T{\simeq}T_c$ 
very small. Any further suppression of $1/\tau(\omega)$, similar 
to that observed in under- and optimally doped samples, could 
potentially produce only a weak feature that would be difficult
to detect experimentally.

To conclude this sub-subsection, we make a comparison between our 
data on the temperature/frequency dependence of the scattering rate 
with earlier results. In the optimally doped Y123 and Bi2212 samples, 
microwave and infrared experiments demonstrated that 
$1/\tau(\omega\rightarrow 0)$ drops abruptly below the 
superconducting transition temperature.\cite{bonn92,romero92} 
A suppression of the scattering rate in the superconducting 
state was confirmed through transport experiments.\cite{salamon} 
These results are consistent with the behavior of $1/\tau(\omega)$ 
plotted in Fig.~\ref{opt-tau}.

In the underdoped regime, the suppression of the scattering rate 
occurs already in the normal state and thus a comparison can be made 
with dc resistivity data. In underdoped cuprates the resistivity is a 
linear function of T for $T>T^*$, but it shows a crossover to a 
steeper slope at $T<T^*$.\cite{ito93} Since dc resistivity is, 
within a constant factor, the zero frequency limit of 
$1/\tau(\omega)$, the crossover behavior could be completely 
accounted by with the low-frequency suppression of the scattering 
rate. We also note that the dc resistivity of underdoped cuprates, 
at least below 300~K, is determined by the charge dynamics in a 
relatively small energy range (below the threshold structure) while 
in the strongly overdoped cuprates, much larger energies are 
involved. It is not quite clear, however, how the $1/\tau(\omega)$ 
spectra in the underdoped cuprates will evolve above room 
temperatures where dc $\rho_{ab}(T)$ is still increasing with 
temperature. In particular, it is not clear whether the 
$1/\tau(\omega)$ will remain linear and temperature-independent at 
high frequencies. 

\subsection{Theoretical models for $1/\tau(\omega)$}

There is yet to be a clearly superior theoretical explanation for 
the peculiar behavior of the infrared optical response presented in 
the previous section, but a few models deserve mentioning. We 
will start here with the models that rely on inelastic scattering 
processes as the mechanism that determines the frequency and 
temperature behavior of the real and imaginary parts of the memory 
function, and will continue with other models later. 

\begin{figure}[htpb]
\leavevmode
\epsfxsize=0.50\columnwidth
\centerline{\epsffile{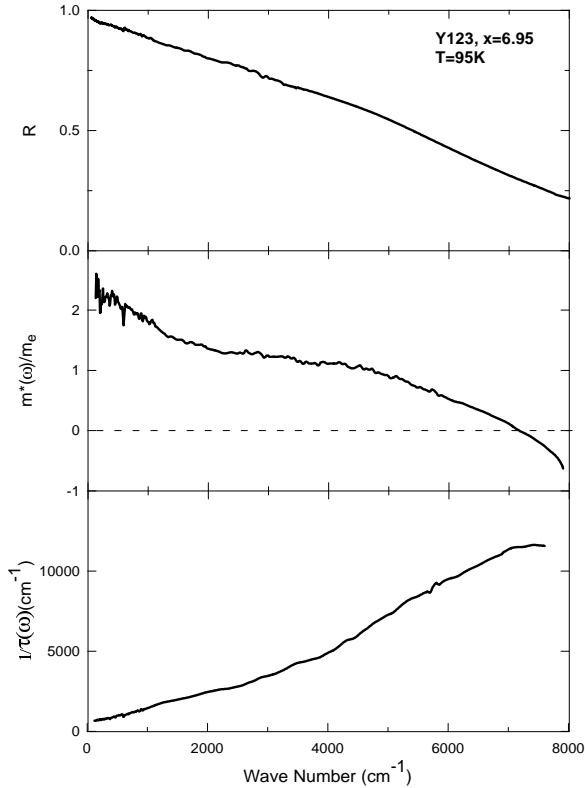}
 }
 \caption{The reflectivity (top panel), the effective mass (middle 
panel), and the frequency dependent scattering rate (lower panel) for 
an optimally doped Y123 crystal shown over a wider frequency range. 
Temperature is 95~K. The effective mass becomes negative for 
$\omega>6000$~cm$^{-1}$ suggesting a breakdown of the validity of 
the single-component approach due to onset of an interband 
transition.}
\label{8000}
\end{figure}

As we have stressed previously, the modeling of the real part of
$M(\omega)$ in terms of the carrier-scattering only makes 
sense if there are reasons to believe that the optical response 
in the energy region under study is predominantly due to mobile 
carriers (no interband contribution) and that there is only one type 
of carrier participating in optical excitations (one-component model).
It is not at all clear that these requirements are satisfied in the 
HTSC cuprates at all frequencies, particularly in the midinfrared 
range, where some of the interband transition processes may have energies 
comparable with the those of the intraband excitations. The situation 
is complicated further by the fact that these contributions do not 
have characteristic sharp features which would make facilitate their 
separation. As an example, a typical frequency dependence of 
the room-temperature ab-plane complex memory function $M(\omega)$, in 
the functional form of $1/\tau(\omega)$ and $m^*(\omega)$, is shown 
in Fig.~\ref{8000} on a large frequency scale for Y123 ($x=6.95$) 
material ($E{\parallel}a$). Evidence for the interband process comes 
from, for example, $m^*(\omega)$ being negative\cite{interband-note} 
at ${\omega}>8000$~cm$^{-1}$.

Nevertheless, there are reasons to believe that the 
carrier-scattering approach can be used at frequencies below 
2000-3000~cm$^{-1}$ where we have presented data in Section~IV. 
First, the conductivity is observed to be temperature 
dependent\cite{footnoteromero} at $\omega<2000-3000$ cm$^{-1}$, and 
it is natural to assign the temperature- dependent part to the "free" 
carrier contribution; Second, as it was noted earlier by Thomas 
{\it et al.},\cite{thomas88a} the number of carriers that one 
obtains using the sum rule analysis for the real part of optical 
conductivity is consistent with estimates from chemical valence 
arguments for the carrier density provided, the sum rule is taken up 
to about 8000~cm$^{-1}$. 

Therefore the carrier-scattering mechanisms is at least a plausible 
mechanism for the optical response in HTSC at frequencies 
less than 2000-3000~cm$^{-1}$. Below we will outline some approaches 
that are based on carrier-scattering mechanisms as well as some 
problems associated with them. 

The first approach is electron-phonon scattering.\cite{Shulga}
While this model qualitatively
reproduces the gap-like depression in $1/\tau(\omega)$ 
at low temperatures (see, for example, calculations presented in 
Figs.~\ref{theory1},\ref{theory2}), it is not nearly as sharp as 
that seen in the experimental data. An even more severe problem is 
the absence of the predicted temperature dependence of 
$1/\tau(\omega)$ at high frequencies. A signature of the 
electron-phonon theories is their prediction of significant 
temperature-induced changes (proportional to $k_B$T at high 
temperatures). Furthermore, as discussed in Section III, within the 
electron-boson scattering scenario, the characteristic temperature 
below which a low-frequency depression in $1/\tau(\omega)$ occurs is 
determined by the high-energy cut-off of the bosonic spectrum ${\cal 
A}_{tr}(\omega)$. The experimental fact is that the characteristic 
temperature in the cuprates, $T^*$, depends on doping level. This is 
inconsistent with electron-phonon scenario, since the phonon cut-off 
is doping-independent. Thus we believe that the electron-phonon 
scattering model fails to reproduce the essential features of the 
experimental data for underdoped cuprates.

It is still possible, however, that phonons play some role in the 
mechanism responsible for the experimentally observed behavior of 
$1/\tau(\omega)$, but in a more unconventional way. We note in this 
respect that the frequency scale in the spectra of $1/\tau(\omega)$ 
associated with the pseudogap state, which does not significantly 
change with doping, remarkably coincidences with the 
high-frequency cut-off energy of the phonon density of states in 
HTSC.\cite{renker88} 

More generally, a serious defect of all models that employ scattering 
of electrons by bosonic excitations to describe the optical response 
of underdoped HTSC is their failure to account for the observed 
behavior in the high-frequency part of $1/\tau(\omega)$ spectra. As 
discussed in Section~IV, underdoped cuprates do not show any 
temperature dependence in $1/\tau(\omega)$ at $\omega>700-800$ 
cm$^{-1}$. On the other hand, in Section~III we saw that scattering 
of electrons by any temperature-independent bosonic spectrum leads to 
a strong temperature dependence of $1/\tau(\omega)$ at high 
frequencies. The only way to get around this contradiction is to 
assume that the boson spectral function ${\cal A}_{tr}(\omega)$ is 
also a function of {\it temperature}: ${\cal A}_{tr}(\omega,T)$. In 
this case, if the absolute value of ${\cal A}_{tr}(\omega,T)$ 
scales properly with temperature, it may account for the observed 
temperature-independent scattering rate at high frequencies. The 
phonon density of states does not show any such 
changes.\cite{renker88} 

One of the mechanisms that may yield a temperature-dependent ${\cal 
A}(\omega)$ function is the scattering of charge carriers on local 
fluctuations towards an antiferromagnetic order. The energy scale 
associated with spin fluctuations is measured\cite{bourges95} to be 
of the order of 50~meV. The features in the scattering rate spectra 
that we observe in the pseudogap state are on the same energy scale, 
supporting such models. This mechanism would also provide a qualitative 
explanation for the doping dependence of the pseudogap. 

\begin{figure}[htpb]
\leavevmode
\epsfxsize=0.6\columnwidth
\centerline{\epsffile{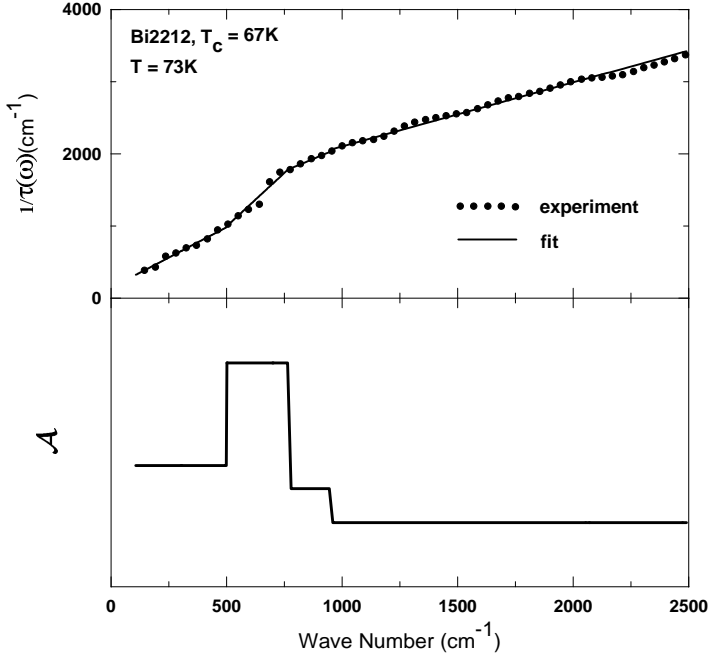}
 }
 \caption{Top panel: scattering rate above $T_c$ in the pseudogap 
state is approximated by straight line segments. Bottom panel: 
slopes of the straight segments are plotted as a function of 
frequency.}
\label{alpha}
\end{figure}

Finally, we can roughly estimate the boson spectral function that is 
needed to obtain the threshold structure in 
$1/\tau(\omega)$ in the pseudogap state at $T<T^*$. This estimate 
can be obtained by inverting the lowest temperature normal-state 
experimental results for $1/\tau(\omega)$ using Allen's expression 
(Eq.~14). The complete inversion formula can be written as ${\cal 
A}(\omega)=1/{\omega}d/d{\omega}[{\omega^2}d/d{\omega}(1/{\tau}(\omega))]$.
\cite{marsiglio96}
Since the process of numerical differentiating greatly amplifies the 
noise level of our spectra, we have chosen the following approach 
to minimize the added noise: The experimentally obtained 
$1/\tau(\omega)$ for underdoped Bi2212 was fitted with four straight 
lines, as shown in Fig.~\ref{alpha} and then the inversion formula was 
applied to the resulting artificial spectrum composed of the straight 
pieces. In this scheme, the exact inversion formula reduces to the 
first derivative, that is the slope of the straight lines. The 
resulting ${\cal A}_{tr}(\omega)$ spectrum is shown in the bottom panel of 
Fig.~\ref{alpha}. Obviously the rather crude approximation of the 
experimental curve prevents us from observing any fine details of the 
spectrum. The significant result is, however, that an intense peak 
in ${\cal A}_{tr}(\omega)$ at 500-700~cm$^{-1}$, superimposed on a broad 
background, is needed to account for the behavior of the scattering 
rate in the pseudogap state if one adopts an electron-boson 
scattering model. 

We note that at least some of the current electron-electron 
scattering models suffer the same problems as the electron-boson 
ones. Namely, they cannot account for the weak or 
completely absent temperature dependence of $1/\tau(\omega)$ spectra 
in underdoped cuprates at high frequencies. In the conventional Fermi 
liquid theory, for example, electron-electron scattering rate is 
proportional to $(\hbar\omega)^2+({\pi}k_BT)^2$, that 
is the frequency and temperature dependence of the scattering rate 
"mirror" one another.\cite{pines66} Another example of this type of 
mirroring is provided by the heavy fermion compound 
URu$_2$Si$_2$\cite{bonnurusi} or the perovskite 
Sr$_2$RuO$_4$.\cite{katsufuji96} In both cases a scaling of 
$\hbar\omega=2k_BT$ collapses the dc resistivity curve on the 
frequency dependent scattering rate curve. This is in contrast to the 
experimental observations of underdoped cuprates where a significant 
frequency dependence, but no temperature dependence, were observed at 
frequencies above 1000~cm$^{-1}$. Other Fermi-liquid type models, 
such as the nested Fermi-liquid (NFL) model or the marginal 
Fermi-liquid (MFL) model, also predict a significant temperature 
dependence at high frequencies.\cite{rieck95,MFL} For example, the 
main assumption of the phenomenological MFL model is that the 
scattering rate varies as $1/\tau(\omega,T)=\alpha\omega + \beta T$ 
where $\alpha$ and $\beta$ are constants of the order of unity. It is 
clear that in the underdoped materials $1/\tau$ does not follow this 
behavior since $\beta=0$, {\it i.e.} there is no temperature 
dependence associated with the linear in frequency scattering rate. 
As we have seen, a temperature dependence of the scattering rate does 
develop, but only in optimally doped and overdoped materials. 

We note that from a completely different point of view, the 
two component model of optical conductivity\cite{tanner92a} where 
the infrared conductivity is divided into a free-carrier and a 
midinfrared component, these observations imply that the mid infrared 
component is temperature independent in underdoped materials.

Some hints regarding the microscopic origin of the scattering 
mechanism in HTSC can be obtained from the analysis of impurity 
effects. As was shown in section IV, the effect of Zn doping was not 
just the introduction of an additional frequency independent term in 
$1/\tau(\omega)$ spectra. Zinc substitution also modifies the 
frequency dependence of the scattering rate suggesting that {\it 
inelastic} processes are affected by Zn as well. The $c$-axis results 
obtained for the sample with 0.425~$\%$ Zn reveal a {\it complete 
suppression} of the pseudogap.\cite{basov96c} The effect of Zn on the 
$c$-axis pseudogap is similar to the one observed in the spin-lattice 
relaxation time 1/$TT_1$.\cite{ishida} The similar concentrations of 
Zn in ceramic pellets of Y124 completely suppress the pseudogap 
feature in the temperature dependence of $1/TT_1$, despite the fact 
that behavior of the Knight shift remains unchanged from that of a 
pure sample. The results obtained from crystals with Zn substitution 
strongly suggest that spin fluctuations may be involved or may even 
be the dominant mechanism of scattering in underdoped cuprates. 

There are several other theoretical models that attempt to explain 
the pseudogap phenomenon from different assumptions. 

The model due to Emery and Kivelson\cite{emerynat} predicts that the 
low carrier density in the underdoped regime may result in 
pairing without pair-pair coherence at temperatures 
well above the actual $T_c$, thus producing a pseudogap. As the
temperature is lowered the phase coherence is established, leading to 
bulk superconductivity. This model would provide an explanation for 
the lack of dramatic changes upon crossing into the superconducting 
state, which is consistent with our optical experiments as well as
ARPES measurements. However, it is not quite clear why the 
high-frequency onset energy of the optical pseudogap does not change 
as a function of doping while $T^*$ and $T_c$ do.

In the spin-charge separation picture,\cite{anderson87} spin 
singlets form at $T^*$, giving rise to a spin gap while the charge 
carriers, holes which are bosons, Bose condense at the 
superconducting transition.\cite{Kotliar88,Rice92,Fukuyama92,Lee92} 
Other models invoke a spin density 
wave\cite{kampf90,barzykin95} in the context of a normal Fermi liquid to 
form a gap in the spin excitations which are the predominant 
scatterers of the charge carriers. 

\subsection{Effect of Zn doping} 

As a small fraction of Cu is substituted by Zn, the optical response 
of underdoped Y124 changes dramatically. Contrary to what is 
seen in the underdoped samples in the pseudogap state or the 
optimally doped sample in the superconducting state, there is no 
structure in the scattering rate neither above nor below $T_c$. 
These results suggest the temperature characterizing the pseudogap 
state $T^*$ can be very small and definitely much lower than 
$T_c$. We note that a similar effect may be seen in overdoped 
cuprates. Therefore we suggest that the addition of Zn may have an 
effect similar to the overdoping of Y124 compounds. 

The normal-state plasma frequency is not affected by Zn 
substitution. Indeed, as was demonstrated by Puchkov {\it et 
al.},\cite{puchkov96a} overdoping does not lead to any significant 
changes in the value of the in-plane plasma frequency. At the 
same time, overdoped compounds usually show a higher dc conductivity 
than their under- or optimally doped counterparts. That is not the 
case in the Y124 material with Zn where the conductivity 
$\sigma_a(\omega\rightarrow 0)$ is reduced. However this may simply 
reflect the fact that Zn is put directly in the CuO$_2$ planes and 
this inevitably causes additional impurity scattering. The latter 
effect is so strong that crystals with 1.25$\%$ of Zn show an 
evidence for a charge-carriers localization behavior in the optical 
conductivity. This point shall be addressed in detail elsewhere.

It is critical to determine whether the effects observed in the Y124 
crystals with Zn substitution are unique for this specific impurity 
or whether other types of disorder would produce similar results.

\subsection{The relation between ab-plane and c-axis pseudogap.}

A comparison of $a$-axis results for Y123 materials with earlier 
$c$-axis data\cite{homes93,basov94c} suggests that the pseudogap 
directly observed in the $c$-axis conductivity at $T<T^*$ is 
necessarily accompanied by a suppression of the in-plane 
$1/\tau(\omega)$ at low frequencies. Indeed, the threshold feature 
in $1/\tau_{a}(\omega)$ is found in underdoped 
crystals at $T<T^*$ {\it only} when the spectrum of 
$\sigma_c(\omega)$ exhibits a pseudogap. The suppression of the 
pseudogap in $\sigma_c(\omega)$, either by an increase of 
temperature above $T^*$, or by an increase of the carrier density 
from x=6.6 to x=6.95 in Y123, or by the substitution of Cu with Zn in 
underdoped Y124,\cite{basov96c} restores the nearly-linear frequency 
dependence of the in-plane scattering rate. Therefore, we conclude 
that the same microscopic mechanism leads to the opening of the 
pseudogap in the interplane response of YBCO crystals {\it and} the 
low-frequency anomalies in the lifetime effects within the CuO$_2$ 
planes. 

It is interesting to note that the frequency shape of $c$-axis 
conductivity is somewhat similar to the bosonic spectral function, 
shown in Fig.~\ref{alpha}, that is needed to reproduce the {\it 
in-plane} behavior of $1/\tau(\omega)$. This suggests that there may 
be an intricate connection between the two.

\subsection{The superconducting state}

One of the most striking features of the curves, in our view, is how 
closely the $1/\tau$ curves for the underdoped cuprates in the 
superconducting state resemble those in the pseudogap state. It is 
useful to compare the energy scales for the various experiments that 
reveal the presence of a pseudogap. 

The maximum gap seen in ARPES experiment is about 
$2\Delta=360$~cm$^{-1}$ (45 meV) whereas the c-axis conductivity (in 
YBCO) shows an onset at ${\sim}200~$cm$^{-1}$ (25 meV) rising to a 
plateau at ~360~cm$^{-1}$ (45~meV). The ab plane $1/\tau$ scale 
is considerably higher with the steepest part of the curve at 
${\sim}500~$cm$^{-1}$ (62~meV) merging with the high frequency linear 
curve around 750~cm$^{-1}$ (93 meV) in all of the materials studied. 
Another high energy scale is the energy range of 
the depression of c-axis conductivity at 
the superconducting transition --- of the order of or larger than 
600~cm$^{-1}$. 

Thus it appears to us that the energy scales associated with the 
pseudogap and with the superconducting state are different. 
In Y124 crystals with Zn substitution, superconductivity persists 
while the pseudogap is suppressed. We also note that in all samples 
we find finite absorption extending down to the lowest frequencies. 
In an $s$-wave superconductivity scenario, this absorption implies a 
very anisotropic superconducting gap. As for the $d$-wave gap models, 
our data may be consistent with the theoretical 
predictions\cite{carbotte95,carbotte96,hirshfeld96} only if one 
assumes a significant amount of impurities present in the crystals. 
This assumption is, however, inconsistent with the linear penetration 
depth observed in the high quality YBCO crystals used in this 
work.\cite{hardy,zhang}

Although changes in $1/\tau(\omega)$ upon crossing into the 
superconducting regime in the optimally doped cuprates are apparently 
dramatic, it may simply be due to the simultaneous formation of the 
pseudogap and superconducting condensate. Also, as noted above, it is 
only in the c-axis conductivity where we see evidence of a larger 
energy scale associated with the superconducting 
state.\cite{basov94c}

In superconducting state, the spectra of the effective mass are 
remarkably similar in all crystals we have studied. In particular, 
the absolute value of $m^*(\omega\rightarrow 0)$ is about 4 both in 
Y123 and Y124 materials. As noted in section III, the zero frequency 
extrapolation of the effective mass gives a square of a ratio of the 
total plasma frequency, $\omega_p$, to the plasma frequency of the 
superconducting condensate, $\omega_{ps}$. This value is in good 
agreement with the results obtained directly from the use of the sum 
rule analysis of $\sigma_1(\omega)$ or from an analysis of the 
imaginary part of the conductivity. The fact that the 
zero-frequency extrapolations of the effective mass are roughly the 
same, $m^*(\omega\rightarrow 0){\simeq}3.5-4$, for all underdoped 
materials suggests that the superfluid condensate density scales with 
the total carrier density in the underdoped cuprates. Therefore, we 
conclude that there are no pairbreaking effects in the pseudogap 
state. However, as doping is increased above optimal, the mass 
enhancement becomes weaker, which indicates a decrease in the 
superfluid density. This behavior is in agreement with the earlier 
$\mu$SR results.\cite{uemura89,uemura91,niedermayer93}

\subsection{The phase diagram and the comparison with c-axis data}

In Fig.~\ref{phase-dia1} we showed a phase diagram where the characteristic 
temperatures $T^*$ (determined from the $c$-axis conductivity) and 
$T_c$ for several different samples from the YBCO family are plotted 
as a function of superfluid density $\omega_{ps}^2 = n_s/m^*$ in the 
CuO$_2$ planes. The superfluid density is obtained from the optical 
conductivity as described in section III. Our choice of superfluid 
density rather than $\omega_p^2$ of the normal-state carriers is 
governed by the fact that the former quantity could be determined 
unambiguously from the real part of the ab plane infrared 
conductivity. From under- to optimally-doped regimes the critical 
temperature scales with the superfluid density in the CuO$_2$ planes and 
the $T_c$ points for YBa$_2$Cu$_3$O$_{6.6}$ and YBa$_2$Cu$_3$O$_{6.95}$ 
crystals fall on the universal dependence first proposed by Uemura {\it et 
al.}\cite{uemura89,uemura91} The $T_c$ of Y124 is 20~$\%$ above the 
universal line. The $T_c$ {\it v.s.} $n_s/m^*$ boundary in the phase 
diagram is well defined since both $T_c$ and $n_s/m^*$ could be determined 
very accurately. However, the $T^*$ {\it v.s.} $n_s/m^*$ 
boundary cannot be determined with same high precision since the
uncertainty in $T^*$ is about 20-30~K, based on the 
$c$-axis conductivity data. The four $T^*$ points correspond to the 
following crystals: $T^*=300$~K - Y123 crystal with $x=6.6$, 
$T^*=180$ K - Y123 with $x=6.7$,\cite{homes93a} $T^*=140$~K - Y124 
crystal and for Y123 with $x=6.95$ $T^*=T_c=93.5$~K. The variation 
of $T^*$ between the different YBCO samples significantly exceeds the 
error in the absolute value of $T^*$. So far crystals with the oxygen 
content less than $x=6.5$ have not been investigated in detail. Thus 
it is unclear if the pseudogap temperature continues to grow as one 
approaches the insulating region in the phase diagram or if it 
saturates at the level of 300-400~K. 

\subsection{Open questions}

At the time of writing this survey of the $ab$-plane 
pseudogap phenomenon, there remain many open questions. 
The first question that must be addressed is whether or not the 
pseudogap state is generic among all high-$T_c$ materials. In 
particular, does anything similar exist for non-cuprate 
superconductors such as BKBO or RENi$_2$B$_2$C. 

It has been suggested that the pseudogap is a manifestation of 
interlayer coupling and is specific to the double layer materials 
such as YBCO and Bi2212.\cite{millis90} Support for this view 
comes from NMR measurements, which show a rather weak depression of 
magnetic susceptibility in La214 in the temperature range where the 
transport data show evidence of a strong suppression of scattering. 
As we have seen from our presentation of the data for overdoped 
single-plane Tl2201 samples, the $ab$-plane $1/\tau(\omega)$ curves 
look similar to those of the two-plane materials {\it in the 
superconducting state}. 

Experimental optical data exists for the one plane La214 material 
\cite{gao93,uchida96,startseva96} which, in the underdoped regime, 
shows a very strong depression in $1/\tau(\omega)$ at low 
temperatures which is consistent with the pseudogap picture. 
However, one must be cautious at this stage since the data from 
various laboratories show considerable variation in the magnitude of 
the effect. In some cases, the structure in reflectivity is so strong 
that it produces an unphysical singularity in the $1/\tau(\omega)$ 
curves.\cite{uchida96} More work on a range of samples must be done 
for this system. Similar strong features are seen in the 
electron-doped Nd$_{2-x}$Ce$_x$CuO$_4$ material.\cite{startseva96} 

It has been suggested that the one-component model of 
 charge transport in the cuprates is particularly unsuited for the 
La214 system where, at least at low doping levels, $\sigma(\omega)$ 
shows a separate midinfrared band\cite{cooper94} rather than a smooth 
free-carrier band with excess conductivity at high frequencies. 
It is also known that at very low doping levels, in the insulating 
state, there is a separate band or several bands 
\cite{thomas91,perkins93} and a one component picture is clearly 
inappropriate. 

Another important effect that needs to be examined is the role of 
impurities. We have seen that Zn has the effect of destroying the 
pseudogap in Y124, both in the $c$-axis $\sigma_1(\omega)$ and in 
the $ab$-plane $1/\tau(\omega)$ curves which acquire a frequency 
dependence similar to what is seen in the overdoped materials. Zn is 
an impurity that has a strong effect on $T_c$ and a systematic study 
of the influence of Zn may help us to isolate its effect on $T^*$, 
the onset of the pseudogap phase, and $T_c$, the onset of 
superconductivity.

Phonons play an important, if perhaps subsidiary, role in high 
temperature superconductivity. As we have seen in section III, the 
standard electron-phonon mechanism predicts a temperature dependent 
$1/\tau(\omega)$ at all frequencies whereas the observations in the 
pseudogap state show a temperature independent 
high-frequency $1/\tau(\omega)$. On the other hand, the frequency 
range of the steepest rise of $1/\tau(\omega)$ falls in the oxygen 
mode region of the phonon spectrum and seems to vary little with 
temperature, chemical composition or doping. This inertness of the 
pseudogap frequency suggests that phonons may be involved in some 
indirect way. 

One process that affects the $ab$-plane conductivity in all 
high-$T_c$ materials is the coupling of the $ab$-plane electrodynamic 
response to $c$-axis LO phonons.\cite{reedyk92e,timusk91,kostur96} To 
separate this process from other processes, it is necessary to 
measure the in-plane optical response on the $ac$ face of an 
underdoped crystal where the LO$_c$ coupled structure 
vanishes.\cite{reedyk92e} 

The signature of the pseudogap state of YBCO materials is that the
in-plane conductivity is enhanced whereas the interplane conductivity 
is suppressed. It is important to find out whether this is manifested 
by other cuprates.

\section{Conclusions}

In our review of the recent optical data, we see that there is a 
universal depression of the real part of the memory function 
$M'(\omega)$, or $1/\tau(\omega)$, below an energy of the order of 
700-800~cm$^{-1}$ in all underdoped materials below a characteristic 
temperature $T^*$. At the optimal doping level $T_c{\simeq}T^*$ and 
in the strongly overdoped regime the gap-like depression is not seen. 
While the high-frequency $1/\tau(\omega)$ was found to be 
temperature-independent in the underdoped cuprates, an obvious 
temperature dependence is seen in the strongly overdoped cuprates. We 
believe that these optical results add to the growing evidence for 
the existence of a normal state pseudogap in the physical response 
function of the underdoped HTSC.

While intense theoretical work has been done to explain the observed 
phenomenon, none of it has been completely successful. It is 
necessary for any theoretical model to explain not only the formation 
of the gap in the $ab$-plane response, but also a wealth of 
phenomena, such as the $c$-axis transport and the remarkable 
temperature dependencies that are observed, both for the $c$-axis 
pseudogap as well as the $ab$-plane response. 

The authors wish to acknowledge a collaboration with the following 
crystal-growing groups, without whom this work would not be possible: 
The Bi2212 single crystals were grown at Stanford University by 
P.~Fournier and A.~Kapitulnik; the Tl2201 single crystals were grown by 
N.N.~Kolesnikov in the Institute of Solid State Physics, Russia, and 
S.~Doyle and A.M.~Hermann in University of Colorado, Boulder; the 
detwinned Y123 crystals came from our long time collaborators, 
R.~Liang, D.A.~Bonn and W.H.Hardy at the University of British 
Columbia; the Y124 single crystals were grown by Bogdan Dabrowski at the 
Northern Illinois University.

We are very grateful to C.C.~Homes, T.~R\~{o}\~{o}m and T.~Startseva 
for allowing us to use their experimental data in this paper.

The authors also wish to acknowledge illuminating discussions with 
A.A.~Abrikosov, P.B.~Allen, A.J.~Berlinsky, D.A.~Bonn, N.~Bontemps, 
J.P.~Carbotte, J.C.~Cooper, O.V.~Dolgov, R.C.~Dynes, V.J.~Emery, W.N.~Hardy, 
C.C.~Homes, M.~Julien, K.~Kallin, S.A.~Kivelson, R.B.~Laughlin, 
P.A.~Lee, A.G.~Loeser, J.W.~Loram, A.J.~Millis, F.~Marsiglio, D.~Pines, 
M.~Strongin, Z.-X.~Shen, D. van der Marel, Y.~Uemura. This work was 
supported by the Canadian Institute of Advanced Research (CIAR) and 
Natural Sciences and Engineering Research Council of Canada. 



\begin{references}
\bibitem{interband-note} Since an interband transition can be roughly 
modeled by an oscillator centered at $\omega_i$, the imaginary part 
of conductivity is negative below $\omega_i$.
\end{references}
\end{document}